\documentclass[aps,prx,twocolumn,floatfix,eqsecnum,superscriptaddress, longbibliography]{revtex4-1}

\usepackage{graphicx}
\usepackage{amssymb,amsmath,braket,indentfirst,bm,setspace,amsthm,theorem,pifont}
\usepackage[colorlinks=true,linktoc=page,linkcolor=blue,citecolor=magenta]{hyperref}
\usepackage[normalem]{ulem}   
\usepackage{cancel,soul}
\usepackage[pdftex]{color}

\def\:={\,\raisebox{0.85pt}{.}\hspace{-2.78pt}\raisebox{2.85pt}{.}\!\!=\,}
\def\=:{\,=\!\!\raisebox{0.85pt}{.}\hspace{-2.78pt}\raisebox{2.85pt}{.}\,}

\begin{document}
\title{Hierarchical Majoranas in a Programmable Nanowire Network}

\author{Zhi-Cheng~Yang} \affiliation{Physics Department, Boston
University, Boston, Massachusetts 02215, USA}

\author{Thomas~Iadecola} 
\affiliation{Joint Quantum Institute and Condensed Matter Theory Center,
Department of Physics, University of Maryland, College Park, Maryland 20742,
USA}

\author{Claudio~Chamon} \affiliation{Physics Department, Boston
University, Boston, Massachusetts 02215, USA}

\author{Christopher~Mudry} \affiliation{Condensed Matter Theory Group,
Paul Scherrer Institute, CH-5232 Villigen PSI, Switzerland}
  
\date{\today} 
\begin{abstract}
We propose a hierarchical architecture for building
``logical'' Majorana zero modes
using ``physical'' Majorana zero modes at the Y-junctions of
a hexagonal network of semiconductor nanowires.
Each Y-junction contains three ``physical" Majoranas,
which hybridize when placed in close proximity, yielding
a single effective Majorana mode near zero energy.
The hybridization of effective Majorana modes on neighboring
Y-junctions is
controlled by applied gate voltages on the links of the honeycomb
network.  This gives rise to a tunable
tight-binding model of effective Majorana modes.
We show that selecting the gate voltages that generate
a Kekul\'e vortex pattern in the set of hybridization amplitudes
yields an emergent ``logical" Majorana zero mode bound to the vortex core.
The position of a logical Majorana can be tuned adiabatically, \textit{without} moving
any of the ``physical" Majoranas or closing any energy gaps, by programming
the values of the gate voltages to change as functions of time.
A nanowire network supporting multiple such ``logical" Majorana zero modes
provides a physical platform for performing adiabatic non-Abelian braiding operations in
a fully controllable manner.
\end{abstract}
\maketitle

\tableofcontents

\section{Introduction}

Topological qubits offer stronger resistance to decoherence by storing
quantum information non-locally.
This property is a driving motivation behind theoretical studies
of topological quantum computation.%
~\cite{nayak}
Majorana zero modes (MZMs), for instance, make up half-a-qubit,
thereby allowing the coding of qubits
non-locally in two far-away Majoranas. There has been a number
of experimental setups proposed to realize MZMs in condensed matter
systems.%
~\cite{alicea2012, lutchyn2018}
One approach aims at
engineering Hamiltonians with effective $p$-wave superconductivity by
proximitizing an $s$-wave superconductor to a semiconductor
nanowire with strong spin-orbit coupling,%
~\cite{sau, jason, lutchyn,oreg,chung,duckheim,potter}
or a topological insulator.%
~\cite{fu,fu2,cook,sun}
Such hybrid systems typically host MZMs at the
endpoints or boundaries of the system. Recently, the theoretically
predicted quantized zero-bias conductance peak at $2e^{2}/h$ in the
presence of MZMs has been observed in indium antimonide
semiconductor nanowires covered with an aluminium superconducting
shell.%
~\cite{zhang}

\begin{figure*}[!t]
\centering
(a)
\includegraphics[width=.45\textwidth]{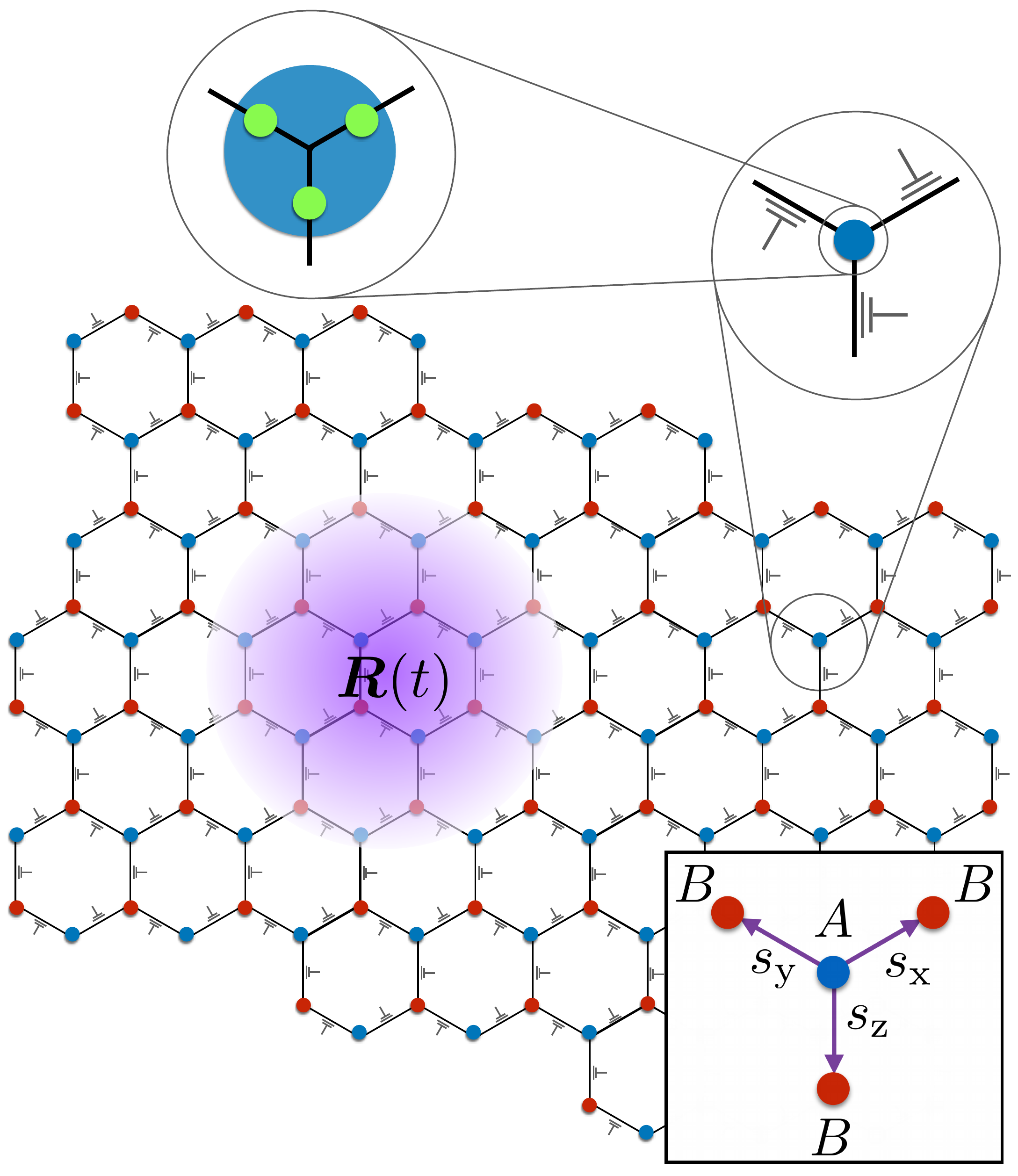}
(b)
\includegraphics[width=.45\textwidth]{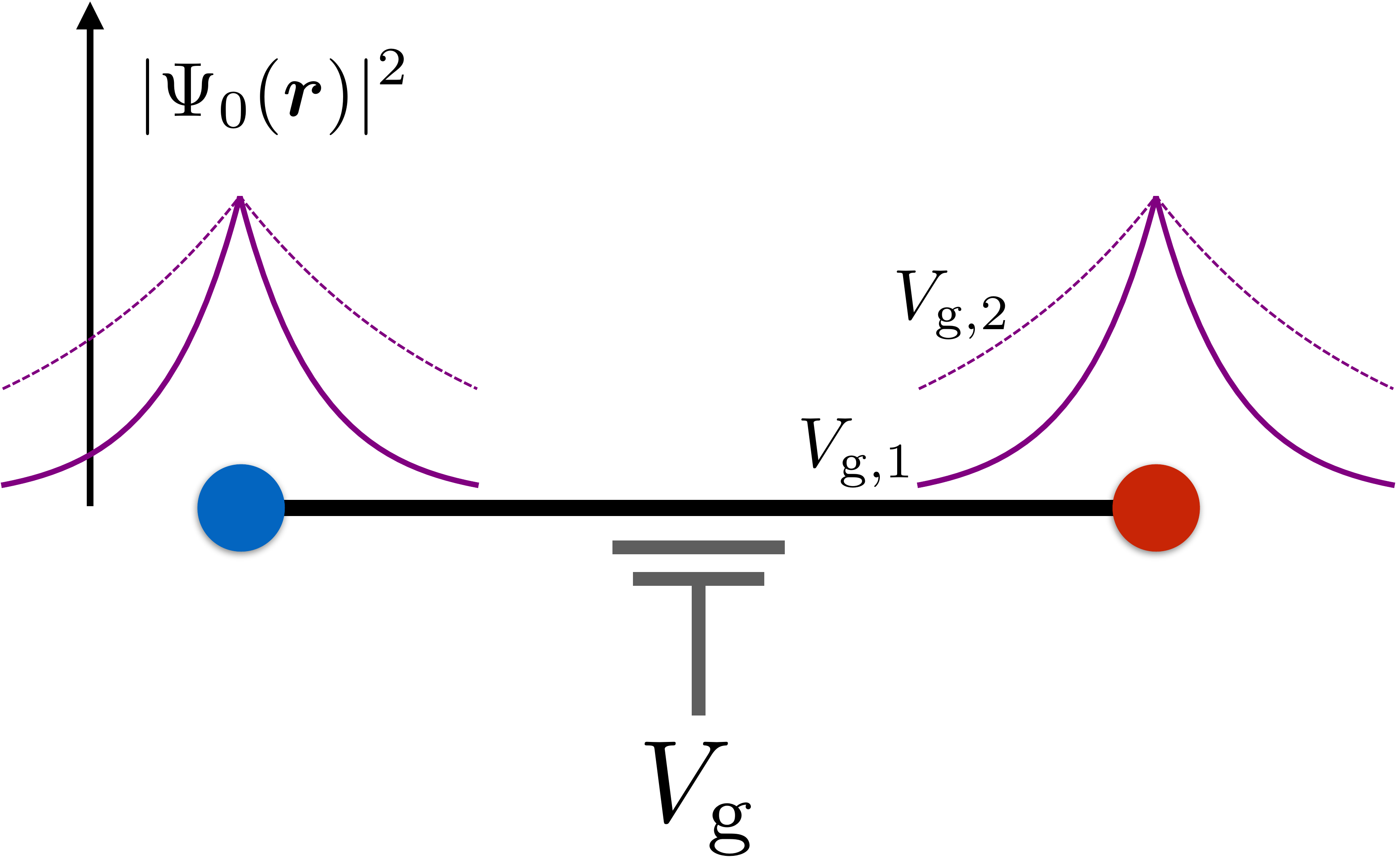}
\caption{
(a)
Hierarchy of Majorana zero modes (MZMs).  We start from an array of
Majorana nanowires, depicted as the black links of a honeycomb
lattice.  Each nanowire furnishes three ``physical" QMZMs (green
circles in the inset) that hybridize locally, leaving one QMZM at each
$Y$-junction (blue circle in the inset).  The resulting effective
QMZMs reside on the sites of a honeycomb lattice (blue and red
circles).  An array of gates (grey ``plungers") provides tunable
hybridization amplitudes for the effective QMZMs.  Writing a
particular pattern of gate voltages gives rise to a Kekul\'e vortex
that binds an emergent ``logical" MZM (purple density profile). The
position $\bm{R}(t)$ of the emergent MZM is arbitrary and can be tuned
continuously as a function of time, so that multiple ``logical" MZMs
can be braided adiabatically.  Inset: Definition of the hexagonal
lattice in terms of the two triangular sublattices $\Lambda^{\,}_{A}$
and $\Lambda^{\,}_{B}$, with the nearest-neighbor vectors $\bm{s}^{\,}_{\alpha}$,
$\alpha=\mathtt{x},\mathtt{y},\mathtt{z}$.
(b)
Controlling the overlap between adjacent effective QMZMs with a gate
voltage $V^{\,}_{\mathrm{g}}$.  The Majorana wavefunctions (purple) decay
exponentially across the length of the nanowire with a decay length
that scales inversely with the topological nanowire gap
$\Delta^{\,}_{\mathrm{nw}}$.
Increasing $V^{\,}_{\mathrm{g}}$ from $V^{\,}_{\mathrm{g},1}$ to
$V^{\,}_{\mathrm{g},2}>V^{\,}_{\mathrm{g},1}$
decreases the nanowire gap, thereby increasing the
wavefunction decay length, and with it the overlap between the two
effective QMZMs (compare solid and dashed curves).
         }
\label{fig:hierarchical} 
\end{figure*}

Despite this progress, there remains the question of how to braid 
MZMs once they are realized experimentally.
For example, many proposals for braiding MZMs involve 
gradually moving microscopic MZMs by applying an array of gates to a single nanowire~\cite{alicea2011}.
There also exists alternative braiding protocols such as coupling to magnetic fluxes~\cite{PhysRevB.88.035121} and measurement-only approaches~\cite{PhysRevB.95.235305}.
In this work, we shall propose a scheme where braiding of MZMs can be implemented
without violating the adiabatic hypothesis. The building blocks of our proposal
are \textit{Majorana nanowires}, i.e.,
semiconductor nanowires supporting Majorana modes bound to
their endpoints at sufficiently low temperatures.
However, the ``logical'' MZMs that are braided are \textit{not}
these elementary Majorana modes residing at the endpoints of the nanowires.
Rather, they are emergent zero modes bound to point topological defects
that can be programmed by gating the nanowires.
These emergent zero modes live in two spatial dimensions,
in contrast to 1D wires where the braiding statistics is intrinsically
ill-defined. The ``logical" MZMs are \textit{hierarchical},
in the sense that they emerge by coupling together a set of Majorana modes
that are themselves the result of the topological state of matter realized
in each nanowire.  The hierarchy of Majorana zero modes that are
used in this work is depicted schematically in Fig.\
\ref{fig:hierarchical}(a).

The hierarchical construction of the ``logical" MZMs
starts from a set of Majorana nanowires.
Since each nanowire is of finite size, the Majorana modes at its
endpoints hybridize weakly and split from
zero energy. We call such a Majorana mode a quasi-Majorana
zero mode (QMZM).
Imagine placing one of the Majorana nanowires on each bond
of a honeycomb lattice.
At each vertex of the honeycomb lattice,
where three nanowires form a Y-junction,
three QMZMs hybridize strongly as their
wavefunctions have large overlaps. This hybridization results
in two QMZMs splitting away from zero energy by an amount
much larger than the energy splitting of the QMZMs bound to the
endpoints of a single nanowire,
leaving a single effective QMZM at each site of the honeycomb lattice.
This is the next level of the hierarchy.
Now, imagine reducing the length of the Majorana nanowires
making up the bonds of the honeycomb lattice.
The increase of the overlap between these effective QMZMs
will then be captured by a tight-binding model for Majorana modes
hopping on the honeycomb lattice. If we assume translation invariant
nearest-neighbor hopping amplitudes, there arises a gapped liquid
with two massive Majorana cones very much as one finds
in Kitaev's honeycomb model in the presence of a magnetic field,%
~\cite{kitaev2006} or in other lattices in the presence of quartic
Majorana interactions.%
~\cite{rahmani2017}

Another gap, which allows for the formation and manipulation of
``logical" MZMs, can then be opened by giving the hopping amplitudes a
Kekul\'e pattern. In practice, this can be done by applying voltages
on the individual Majorana nanowires, which modulates the
hybridization of nearest-neighbor effective QMZMs. To see how, recall
that the topological gap $\Delta^{\,}_{\mathrm{nw}}$ in a Majorana nanowire
\textit{decreases} when a gate voltage $V^{\,}_{\mathrm{g}}$ is applied,
thereby increasing the hybridization.%
~\cite{sau,jason,lutchyn,oreg}
Decreasing the size of the topological gap \textit{increases} the
decay length of the QMZMs, thereby increasing the overlap of their
wavefunctions, see Fig.\ \ref{fig:hierarchical}(b). Thus, by programming
the set of gate voltages applied to every bond, one can exercise
control over every hopping amplitude in the effective tight-binding
model.

Furthermore, one can program these hopping amplitudes in a
position-dependent manner so as to ``write'' an arbitrary number $v$
of Kekul\'e vortices into the system.  This is achieved by modulating
the gate voltages as
$V^{\,}_{\mathrm{g}}\to V^{\,}_{\mathrm{g}}+\delta V^{\,}_{\mathrm{g}\,\bm{r},\alpha}$,
where
\begin{widetext}
\begin{equation}
\delta V^{\,}_{\mathrm{g}\,\bm{r},\alpha}\:=
V^{\,}_{0}\,
\cos
\left(
\bm{K}^{\,}_{+}\cdot\bm{s}^{\,}_{\alpha}
+
(\bm{K}^{\,}_{+}-\bm{K}^{\,}_{-})\cdot\bm{r}
+
\sum_{n=1}^{v}
q^{\,}_{n}\,
\mathrm{arg}\left(\bm{r}-\bm{R}^{\,}_{n}\right)
\right).
\label{eq:voltage}
\end{equation}
\end{widetext}
Here, $\bm{r}$ is a point in one of the triangular sublattices of the
honeycomb lattice, $\bm{s}^{\,}_{\alpha}$
($\alpha=\mathtt{x},\mathtt{y},\mathtt{z}$) are the nearest-neighbor
vectors connecting to the other sublattice (see Fig.\ \ref{fig:hierarchical}),
$\bm{K}^{\,}_{+}=-\bm{K}^{\,}_{-}$ are the corners of the Brillouin
zone of the honeycomb lattice.  The
vorticities $q^{\,}_{n}=\pm1$ ($n=1,\dots,v$) and positions
$\bm{R}^{\,}_{n}$ are here merely parameters that can be tuned at
will.  Kekul\'e vortices have been shown to bind zero-energy modes in
graphene,%
~\cite{hou2007,hou2008}
analogs of which also appear in photonic crystals.%
~\cite{iadecola}
Similar physics arises here,
with the crucial distinction that the zero modes are now of Majorana
nature, owing to the fact that the underlying tight-binding model is
one of Majoranas. It is the MZMs localized near the core of each
vortex that we shall call the ``logical'' MZMs, which constitute the
final level of the hierarchy.  Because the positions $\bm{R}_{n}$ of
the vortices are merely parameters, they can be tuned simply by
changing the voltages on each wire as a function of time, like
addressing pixels on a screen. Therefore, in a system with multiple
vortices, this scheme would allow one to move and braid the logical
MZMs adiabatically.

The rest of the paper is organized as follows.
We present the realization with Majorana nanowires in Sec.\
\ref{Realization with Majorana nanowires}
of an analogue of a $p+\mathrm{i} p$ superconductor
belonging to the symmetry class D.
We determine the conditions under which the Kekul\'e dimerization
controls the gap. We define a scaling limit that allows one to
derive a simple model of free Majoranas
with nearest-neighbor hopping amplitudes
on a honeycomb lattice in Sec.\ \ref{sec: free majorana}.
In this scaling limit,
the low-energy effective theory has higher symmetry,
belonging to symmetry class BDI.
We explicitly solve for the MZM bound to a Kekul\'e vortex.
We further show that the Kekul\'e vortices indeed have the braiding
statistics of MZMs.
In Sec.\ \ref{sec: Zero modes bound to Kekule ...},
we demonstrate numerically the emergence of an MZM bound
to the core of a Kekul\'e vortex away from the scaling limit.
Section \ref{sec: Transporting logical majoranas} discusses possible
experimental measurement schemes for the emergent MZMs and demonstrates
the feasibility of our setup using realistic experimental parameters.
We conclude with a summary and outlook for future directions in Sec.\ \ref{sec: summary}.

\begin{figure}[t]
\centering
\includegraphics[width=.3\textwidth]{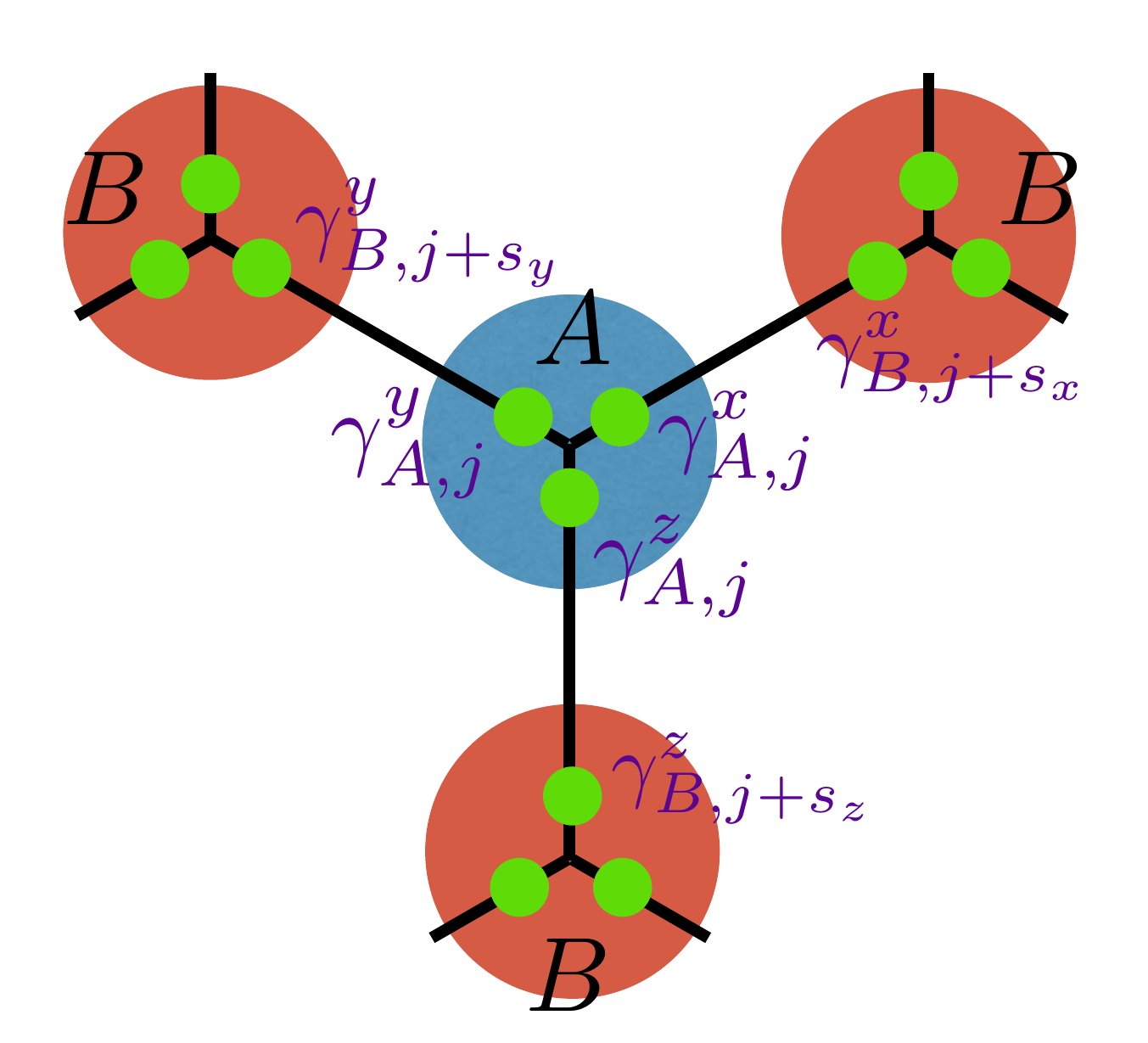}
\caption{
A Y-junction built from Majorana nanowires.
The QMZMs are depicted as green dots.
Effectively, there are three flavors of QMZMs on each
lattice site. We label the operators creating QMZMs by
$\hat{\gamma}^{\alpha}_{S,j}$, where $\alpha=\mathtt{x,y,z}$ denotes the
bond to which the QMZM belongs, while $S=A,B$ denotes the sublattices,
and $j$ is the label for the lattice sites.
        }
\label{fig:junction} 
\end{figure}

\section{Realization with Majorana nanowires}
\label{Realization with Majorana nanowires}

The building block that we shall use in this paper
is a nanowire which at low temperatures
supports a topological superconducting gap
$\Delta^{\,}_{\mathrm{nw}}$.
Because of the topological gap $\Delta^{\,}_{\mathrm{nw}}$,
the nanowire hosts a pair of QMZMs at its endpoints when superconducting.
We shall call such a nanowire a ``Majorana nanowire."

The main idea of this paper is to imagine that each nearest-neighbor
bond of the honeycomb lattice is realized by a Majorana
nanowire. There are two energy scales in the problem: a hybridization
${{U}}$ and a hopping amplitude ${{t}}$, as we now explain.

On the one hand, three Majorana nanowires
must  meet
at the sites of the honeycomb lattice, thereby realizing a Y-junction
of Majorana nanowires, as shown in Fig.~\ref{fig:junction}.
Effectively (i.e., below the energy gap $\Delta^{\,}_{\mathrm{nw}}$  
of an isolated Majorana nanowire),
we have three flavors of QMZMs
on each site of the honeycomb lattice.
The pairwise hybridization
among the three QMZMs
will split their quasidegeneracy by an energy scale $|{{U}}|$.
Then, only one QMZM remains below the energy scale $|{{U}}|$ on any given
Y-junction (site of the honeycomb lattice). Thus, each Y-junction effectively
contributes a single emergent Majorana mode.

On the other hand,
the pair of QMZMs bound to the two ends of a Majorana nanowire
are split away from zero energy by the energy scale ${{t}}$ that results from
the overlap of their wavefunctions. This hybridization increases
as each Majorana nanowire is shortened, inducing a
nearest-neighbor hopping amplitude ${{t}}$ for the three pairs
of QMZMs localized on nearest-neighbor Y-junctions of Majorana nanowires.

Hence, working at energies below the topological gap $\Delta^{\,}_{\mathrm{nw}}$  
of a Majorana nanowire, we have outlined the construction of an effective
six-band tight-binding model on the honeycomb lattice using Majorana nanowires.
Below we shall discuss this construction in more detail.

\begin{figure}[t]
\centering
\includegraphics[width=.4\textwidth]{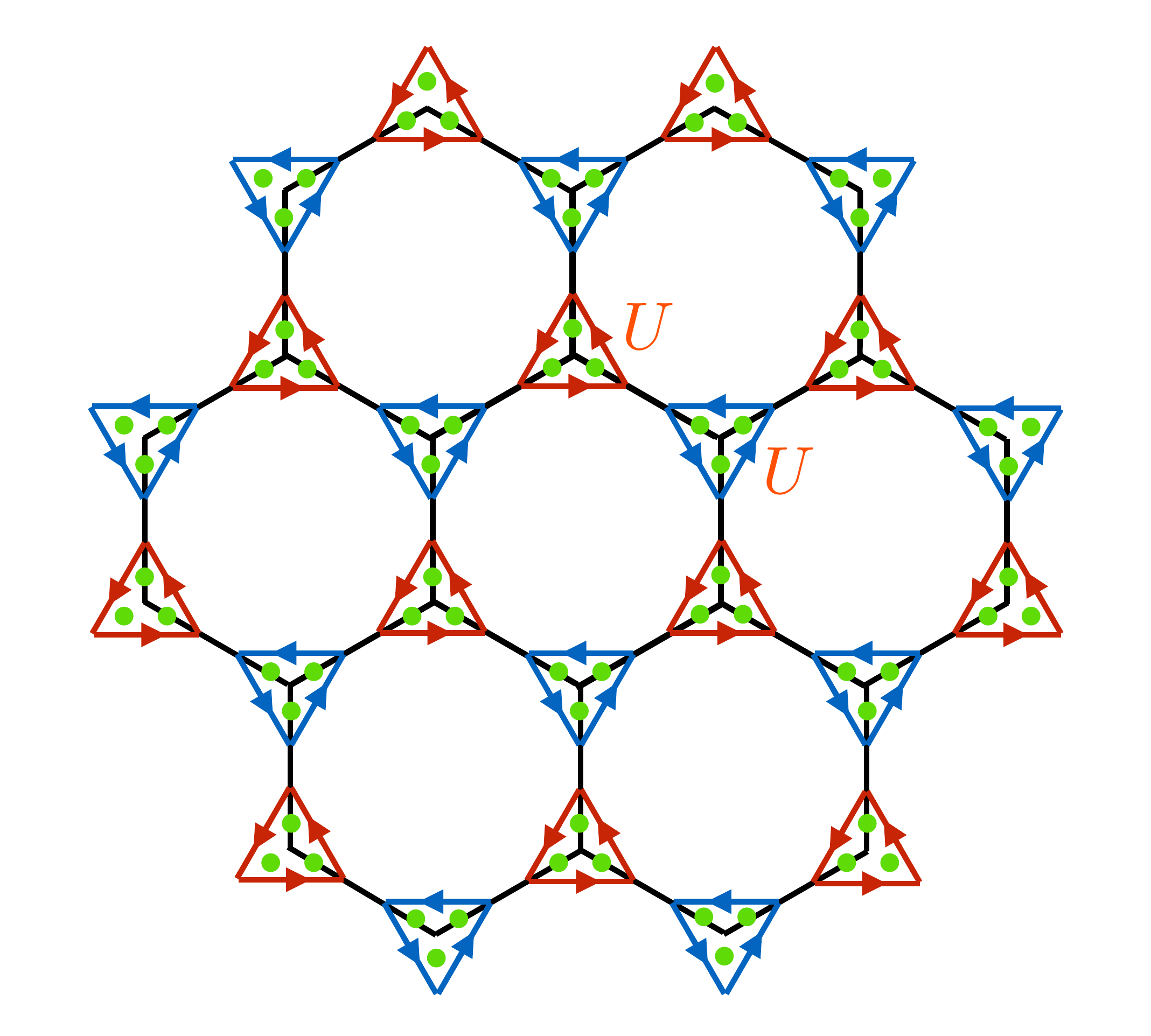}
\caption{
Representation of the trimer limit
defined by the ground state of
Hamiltonian (\ref{eq:trimer}).
The MZMs at each Y-junction are
represented by green dots.
Their pairwise hybridization ${{U}}$ 
is represented by directed bonds
arranged along the edges of a triangle.
The blue and red triangles encircles sites from
sublattices $\Lambda^{\,}_{A}$ and $\Lambda^{\,}_{B}$, respectively.
The hybridization energy scale
for blue and red triangles is ${{U}}$.
The pattern of arrows along the edges of each triangle defines the
order in which two Majorana operators are to be multiplied
with the convention that ${{U}}$ is positive
for this order of multiplication.
         }
\label{fig:trimer} 
\end{figure}

\subsection{Trimer limit (${{U}}\ne 0$, ${{t}}=0$)}

Consider a honeycomb lattice $\Lambda$
made of two interpenetrating triangular lattices
$\Lambda^{\,}_{A}$
and
$\Lambda^{\,}_{B}$.
We shall label the bonds of the honeycomb lattice by
$\alpha=\mathtt{x},\mathtt{y},\mathtt{z}$ depending on their orientations,
as shown in Fig.\ \ref{fig:junction}.
Each bond of the honeycomb lattice realizes a Majorana nanowire.
We shall thus associate to each bond of the honeycomb lattice
a pair of Majorana operators as depicted in
Fig.\ \ref{fig:junction}.
If the label $S=A,B$ distinguishes between the triangular sublattices
$\Lambda^{\,}_{A}$
and
$\Lambda^{\,}_{B}$,
and if the label $j$ stands for a site from $\Lambda^{\,}_{S}$,
then the Majorana algebra reads
\begin{subequations}
\begin{equation}
\left\{
\hat{\gamma}^{\alpha}_{S,j},
\hat{\gamma}^{\alpha'}_{S',j'}
\right\}=
2
\delta^{\,}_{\alpha,\alpha'}
\delta^{\,}_{S,S'}
\delta^{\,}_{j,j'}
\end{equation}
with the Majorana reality condition
\begin{equation}
\hat{\gamma}^{\alpha\dag}_{S,j}=
\hat{\gamma}^{\alpha}_{S,j}.
\end{equation}
\end{subequations}
These Majorana operators stand at the first level of the hierarchy.

The trimer limit occurs for ${{t}}=0$.
The Hamiltonian describing this limit is
\begin{align}
\label{eq:trimer}
\widehat{H}^{\,}_{\mathrm{trimer}}\:=&
\!\!\!
\sum_{S=A,B}
\sum_{j\in\Lambda^{\,}_{S}}
\!
\mathrm{i}
{{U}}
\!
\left(
\hat{\gamma}^{\mathtt{x}}_{S,j}\,
\hat{\gamma}^{\mathtt{y}}_{S,j}
\!+\!
\hat{\gamma}^{\mathtt{y}}_{S,j}\,
\hat{\gamma}^{\mathtt{z}}_{S,j}
\!+\!
\hat{\gamma}^{\mathtt{z}}_{S,j}\,
\hat{\gamma}^{\mathtt{x}}_{S,j}
\right).
\end{align}
We represent in Fig.\ \ref{fig:trimer}
the trimer limit as a decorated honeycomb lattice.
Hybridization within each Y-junction
is represented by a directed arrow relating a pair of MZMs.
The direction of the arrows along the edges of each triangle
defines the order in which two Majorana operators are to be multiplied.
It fixes the sign of the hybridization ${{U}}$ to be positive
along the arrow. Reversing the chirality of the red or blue triangles
thus amounts to reversing the sign of ${{U}}$.

Hamiltonian~(\ref{eq:trimer}) is the sum
over $S=A,B$ and $j\in\Lambda^{\,}_{S}$
of the pairwise commuting operators
\begin{subequations}
\label{eq:trimer solution}
\begin{equation}
\mathrm{i}{{U}}
\left(
\hat{\gamma}^{\mathtt{x}}_{S,j}\,
\hat{\gamma}^{\mathtt{y}}_{S,j}
+
\hat{\gamma}^{\mathtt{y}}_{S,j}\,
\hat{\gamma}^{\mathtt{z}}_{S,j}
+
\hat{\gamma}^{\mathtt{z}}_{S,j}\,
\hat{\gamma}^{\mathtt{x}}_{S,j}
\right).
\label{eq:trimer solution a}
\end{equation}
As each one of these operators has the three single-particle eigenvalues
\begin{equation}
-\sqrt{3}\,{{U}},
\qquad
0,
\qquad
+\sqrt{3}\,{{U}},
\label{eq:trimer solution b}
\end{equation}
with the Majorana zero mode
\begin{equation}
\hat{\eta}\:=
\frac{1}{\sqrt{3}}
\left(
\hat{\gamma}^{\mathtt{x}}_{S,j}
+
\hat{\gamma}^{\mathtt{y}}_{S,j}
+
\hat{\gamma}^{\mathtt{z}}_{S,j}
\right),
\label{eq:trimer solution c}
\end{equation}
\end{subequations}
Hamiltonian~(\ref{eq:trimer})
supports three doubly-degenerate flat bands with the single-particle energies
(\ref{eq:trimer solution b}),
respectively.

\begin{figure}[t]
\centering
\includegraphics[width=.35\textwidth]{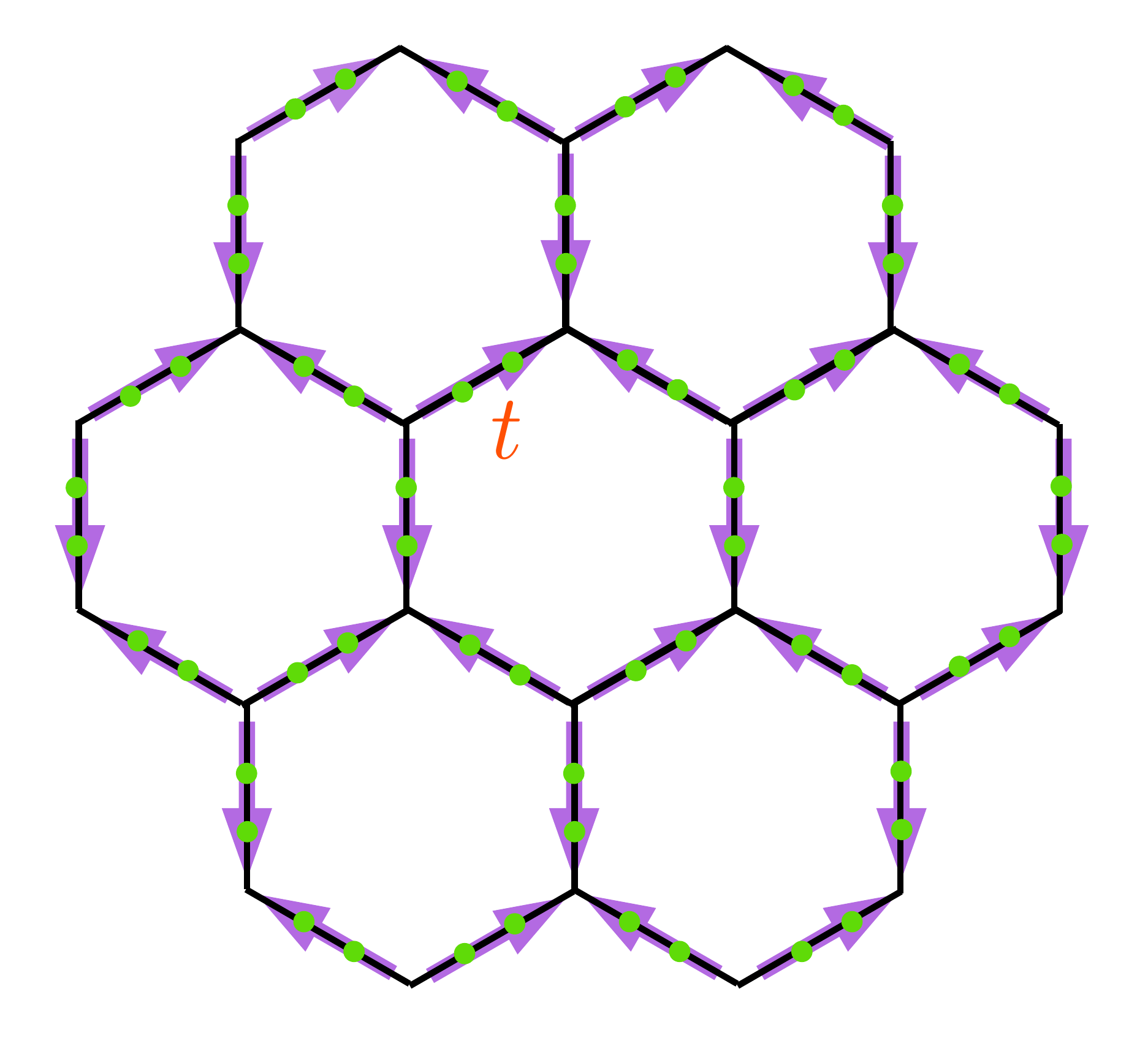}
\caption{
Representation of the dimer limit
defined by the ground state of
Hamiltonian~(\ref{eq:dimer}).
The arrows specify the order in which
Majorana operators (the green dots)
enter Hamiltonian~(\ref{eq:dimer}),
with the convention that operators on sublattice $\Lambda^{\,}_{A}$
are to the left of
operators from sublattice $\Lambda^{\,}_{B}$ along an arrow.
With this convention, the hopping amplitude $t$ is positive along
an arrow.
        }
\label{fig:dimer} 
\end{figure}

\subsection{Dimer limit (${{t}}\ne 0$, ${{U}}=0$)}
\label{subsec: Dimer phase with no zero mode}

The dimer limit occurs for ${{U}}=0$.
The Hamiltonian describing this limit is
\begin{subequations}
\begin{equation}
\widehat{H}^{\,}_{\mathrm{dimer}}\:=
\sum_{j\in\Lambda^{\,}_{A}}
\sum_{\alpha=\mathtt{x},\mathtt{y},\mathtt{z}}
\mathrm{i}{{t}}\,
\hat{\gamma}^{\alpha}_{A,j}\,
\hat{\gamma}^{\alpha}_{B,j+\bm{s}^{\,}_{\alpha}}.
\label{eq:dimer}
\end{equation}
Here, $\bm{s}^{\,}_{\alpha}$
are the unit vectors connecting the three sites in
$\Lambda^{\,}_{B}$ that are nearest-neighbor to a site in
$\Lambda^{\,}_{A}$, i.e.,
\begin{equation}
\bm{s}^{\,}_{\mathtt{z}}\:=
\begin{pmatrix}
0 \\ -1
\end{pmatrix},
\quad
\bm{s}^{\,}_{\mathtt{x}}\:=
\begin{pmatrix}
+\sqrt{3}/2 \\ 1/2
\end{pmatrix},
\quad
\bm{s}^{\,}_{\mathtt{y}}\:=
\begin{pmatrix}
-\sqrt{3}/2 \\ 1/2
\end{pmatrix}.
\label{eq:bond}
\end{equation}
\end{subequations}
One may represent this Hamiltonian
as is done in Fig.\ \ref{fig:dimer}.
The energy scale ${{t}}$ results from the finite lengths of Majorana nanowires,
which allows the pair of wavefunctions of the QMZMs bound to the two ends of
the semiconductor nanowire to have a nonvanishing overlap.
This overlap leads to a splitting of their energies away from $0$
by the amount $\pm|{{t}}|$.

Hamiltonian~(\ref{eq:dimer}) is the sum
over $j\in\Lambda^{\,}_{A}$ and $\alpha=\mathtt{x},\mathtt{y},\mathtt{z}$
of the pairwise commuting operators
\begin{subequations}
\label{eq:dimer solution}
\begin{equation}
\mathrm{i}{{t}}\,
\hat{\gamma}^{\alpha}_{A,j}\,
\hat{\gamma}^{\alpha}_{B,j+\bm{s}^{\,}_{\alpha}}.
\label{eq:dimer solution a}
\end{equation}
As each one of these operators has the two single-particle eigenvalues
\begin{equation}
-|{{t}}|,
\qquad
+|{{t}}|
\label{eq:dimer solution b}
\end{equation}
\end{subequations}
Hamiltonian~(\ref{eq:dimer})
supports two triply-degenerate flat bands with the single-particle energies
(\ref{eq:dimer solution b}),
respectively. The single-particle energies
(\ref{eq:dimer solution b})
correspond to the fermionic state
\begin{equation}
\hat{c}^{\alpha\dag}_{j}|0\rangle\:=
\frac{1}{2}
\left(
\hat{\gamma}^{\alpha}_{A,j}
-
\mathrm{i}\,
\hat{\gamma}^{\alpha}_{B,j+\bm{s}^{\,}_{\alpha}}
\right)\,
|0\rangle,
\qquad
\hat{c}^{\alpha}_{j}|0\rangle\:=0,
\end{equation}
being empty or occupied, respectively.
There is no zero mode in the dimer limit.

\subsection{Reversal of time}
\label{subsec: Reversal of time}

We shall define the action of time reversal by the rules
\begin{equation}
\mathrm{i}\mapsto
-\mathrm{i},
\qquad
\hat{\gamma}^{\alpha}_{A,j}\mapsto
+\hat{\gamma}^{\alpha}_{A,j},
\qquad
\hat{\gamma}^{\alpha}_{B,j+\bm{s}^{\,}_{\alpha}}\mapsto
-\hat{\gamma}^{\alpha}_{B,j+\bm{s}^{\,}_{\alpha}}.
\end{equation}
The motivation for this definition is that we would like to interpret
\begin{equation}
\hat{c}^{\alpha}_{A,j}\:=
\frac{1}{2}
\left(
\hat{\gamma}^{\alpha}_{A,j}
+
\mathrm{i}\,
\hat{\gamma}^{\alpha}_{B,j+\bm{s}^{\,}_{\alpha}}
\right)
\end{equation}
as a fermion operator localized on the directed bond
$\langle j\in\Lambda^{\,}_{A},j+\bm{s}^{\,}_{\alpha}\in\Lambda^{\,}_{B}\rangle$
of the honeycomb lattices that is left invariant by the operation
of time reversal.

One verifies that
Hamiltonian~(\ref{eq:dimer})
is even under reversal of time while
Hamiltonian~(\ref{eq:trimer})
is odd under reversal of time, i.e.,
\begin{equation}
\widehat{H}^{\,}_{\mathrm{dimer}}\mapsto
+\widehat{H}^{\,}_{\mathrm{dimer}},
\qquad
\widehat{H}^{\,}_{\mathrm{trimer}}\mapsto
-\widehat{H}^{\,}_{\mathrm{trimer}}.
\end{equation}
Although $\widehat{H}^{\,}_{\mathrm{trimer}}$ is odd under time reversal,
the zero-energy flat band transforms trivially whereas the finite-energy
bands are interchanged.

\subsection{Hamiltonian for the nanowire network}
\label{subsec: Total Hamiltonian}

When both ${{U}}\neq0$ and ${{t}}\neq0$,
we can write the noninteracting Hamiltonian in momentum space as
\begin{subequations}
\label{eq: def H honeycomb nanonwires}
\begin{equation}
\begin{split}
\widehat{H}^{\,}_{\mathrm{wire}}\:=&\,
\widehat{H}^{\,}_{\mathrm{trimer}}
+
\widehat{H}^{\,}_{\mathrm{dimer}}
=
\int\limits^{\,}_{\Omega^{\mathrm{K}}_{\mathrm{BZ}}}\mathrm{d}^{3}\bm{k}\,
\widehat{\Psi}^{\dag}_{\bm{k}}\,
\mathcal{H}^{\,}_{\mathrm{wire}}\,
\widehat{\Psi}^{\,}_{\bm{k}},
\end{split}
\label{eq:wire}
\end{equation}
with the spinor
\begin{equation}
\widehat{\Psi}^{\dag}_{\bm{k}}=
\begin{pmatrix}
\hat{\gamma}^{\mathtt{x}}_{A,\bm{k}}
&
\hat{\gamma}^{\mathtt{y}}_{A,\bm{k}}
&
\hat{\gamma}^{\mathtt{z}}_{A,\bm{k}}
&
\hat{\gamma}^{\mathtt{x}}_{B,\bm{k}}
&
\hat{\gamma}^{\mathtt{y}}_{B,\bm{k}}
&
\hat{\gamma}^{\mathtt{z}}_{B,\bm{k}}
\end{pmatrix}
\end{equation}
and the single-particle Hamiltonian
\begin{widetext}
\begin{equation}
\mathcal{H}^{\,}_{\mathrm{wire}}=
\begin{pmatrix} 
0
&
+\mathrm{i}{{U}}/2
&
-\mathrm{i}{{U}}/2
&
+\frac{\mathrm{i}{{t}}}{2}\,
e^{\mathrm{i}\bm{k}\cdot\bm{s}^{\,}_{\mathtt{x}}}
&
0
&
0
\\
-\mathrm{i}{{U}}/2
&
0
&
+\mathrm{i}{{U}}/2
&
0
&
+\frac{\mathrm{i}{{t}}}{2}\,
e^{\mathrm{i}\bm{k}\cdot\bm{s}^{\,}_{\mathtt{y}}}
&
0
\\
+\mathrm{i}{{U}}/2
&
-\mathrm{i}{{U}}/2
&
0
&
0
&
0
&
+\frac{\mathrm{i}{{t}}}{2}\,
e^{\mathrm{i} \bm{k}\cdot\bm{s}^{\,}_{\mathtt{z}}}
\\
-\frac{\mathrm{i}{{t}}}{2}\,
e^{-\mathrm{i}\bm{k}\cdot\bm{s}^{\,}_{\mathtt{x}}}
&
0
&
0
&
0
&
+\mathrm{i}{{U}}/2
&
-\mathrm{i}{{U}}/2
\\
0
&
-\frac{\mathrm{i}{{t}}}{2}\,
e^{-\mathrm{i}\bm{k}\cdot\bm{s}^{\,}_{\mathtt{y}}}
&
0
&
-\mathrm{i}{{U}}/2
&
0
&
+\mathrm{i}{{U}}/2
\\
0
&
0
&
-\frac{\mathrm{i}{{t}}}{2}\,
e^{-\mathrm{i}\bm{k}\cdot\bm{s}^{\,}_{\mathtt{z}}}
&
+\mathrm{i}{{U}}/2
&
-\mathrm{i}{{U}}/2
&
0
\end{pmatrix}.
\label{eq:wire k}
\end{equation}
\end{widetext}
\end{subequations}


The single-particle 
Hamiltonian~\eqref{eq:wire k}
is of Bogoliubov-de Gennes (BdG) form.
This is to say that out of its six Majorana bands,
three have positive single-particle energies,
three have negative single single-particle energies,
and there exists an antiunitary transformation such that
the six bands can be organized into three pairs such that
for any one of these three pairs
the Majorana band with positive single-particle energy
maps to the Majorana band with negative single-particle energy
and vice versa under the antiunitary transformation.

When $|{{U}}/{{t}}|\ll1$,
the two flat bands of $\widehat{H}^{\,}_{\mathrm{dimer}}$
acquire a dispersion with a bandwidth that is controlled by
$|{{U}}|$. Both bands are topologially trivial.
We will not consider this limit anymore in the paper.

When $|{{t}}/{{U}}|\ll1$,
the zero-energy modes (\ref{eq:trimer solution c})
of $\widehat{H}^{\,}_{\mathrm{trimer}}$ that are localized on
the sites of the honeycomb lattice get hybridized by
$\widehat{H}^{\,}_{\mathrm{dimer}}$. More precisely,
the twofold degenerate flat band in the Brillouin zone
$\Omega^{\,}_{\mathrm{BZ}}$
arising from the zero mode $\hat{\eta}$
defined in Eq.\ (\ref{eq:trimer solution c}) when ${{t}}/{{U}}=0$
turns into two bands related by particle-hole symmetry.
The bandwidth for this pair of Majorana bands is of order $|{{t}}|$.
These emergent low-energy  Majorana modes realize
the second level of the hierarchy of Majoranas. The limit $|{{U}}|\gg|{{t}}|$
enforces the first hierarchical reduction in the number of
effective Majorana modes. We now turn to a quantitative
analysis of the band structure of the Hamiltonian~\eqref{eq:wire k} 
in this limit.

\begin{figure}[!t]
\centering
\includegraphics[width=0.48\textwidth]{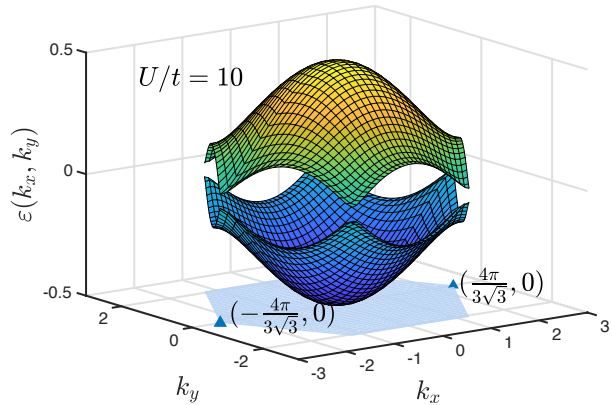}
\caption{
The pair of particle-hole symmetric bands with the lowest energies for
Hamiltonian~(\ref{eq:wire}) when ${{U}}/{{t}}=10$ with ${{U}}>{{t}}>0$.
A Haldane gap appears at the corners of
the Brillouin zone $\Omega^{\,}_{\mathrm{BZ}}$ (depicted in light blue).
The magnitude of the Haldane gap follows from
$\varepsilon^{\,}_{\pm}(\bm{K}^{\,}_{+})=
\varepsilon^{\,}_{\pm}(\bm{K}^{\,}_{-})\approx
\pm\frac{{{t}}^{2}}{2\sqrt{3}\,{{U}}}+\mathcal{O}({{t}}^{4}/{{U}}^{3})$. The energies are plotted in units of $t$.
        }
\label{fig:band haldane unfold} 
\end{figure}

In Fig.~\ref{fig:band haldane unfold},
we plot the band structure for ${{t}}/{{U}}=0.1$ with ${{U}}>{{t}}>0$.
We find that a gap opens at the corners of the Brillouin zone.
We shall call this gap the Haldane gap. This terminology will be explained
when we introduce the single-particle Hamiltonian
(\ref{eq: def Haldane term}) and show that
it opens a spectral gap and endows Majorana bands with non-vanishing
Chern numbers.
A direct calculation of the eigenvalues at $\bm{K}^{\,}_{\pm}$
shows that the energies
of the two bands at $\bm{K}^{\,}_{\pm}$ are given by
\begin{align}
\varepsilon^{\,}_{\pm}(\bm{K}^{\,}_{+})=&\,
\varepsilon^{\,}_{\pm}(\bm{K}^{\,}_{-})
\nonumber\\
=&\,
\pm\frac{1}{4}
\left(
\sqrt{3}\,{{U}}
-
\sqrt{3{{U}}^{2}+4{{t}}^{2}}
\right)
\nonumber\\
\approx&\,
\pm\frac{{{t}}^{2}}{2\sqrt{3}\,{{U}}}
+
\mathcal{O}\left(\frac{{{t}}^{4}}{{{U}}^{3}}\right).
\label{eq: Haldane gap nanowire network}
\end{align}
We thus find that the Haldane gap is of order
${{t}}^{2}/{{U}}$ and, as such, can be explained within second-order
perturbation theory. Upon linearization of the single-particle Hamiltonian
in the vicinity of $\bm{K}^{\,}_{\pm}$,
this gap can be interpreted as a Haldane mass
that implements the microscopic breaking of
time-reversal symmetry.%
~\cite{haldane}.
The counterpart of this phase in the Kitaev's honeycomb model 
is the non-Abelian topologically ordered phase stabilized by a magnetic field.%
~\cite{kitaev2006}

\begin{figure*}[!t]
\centering
\includegraphics[width=1\textwidth]{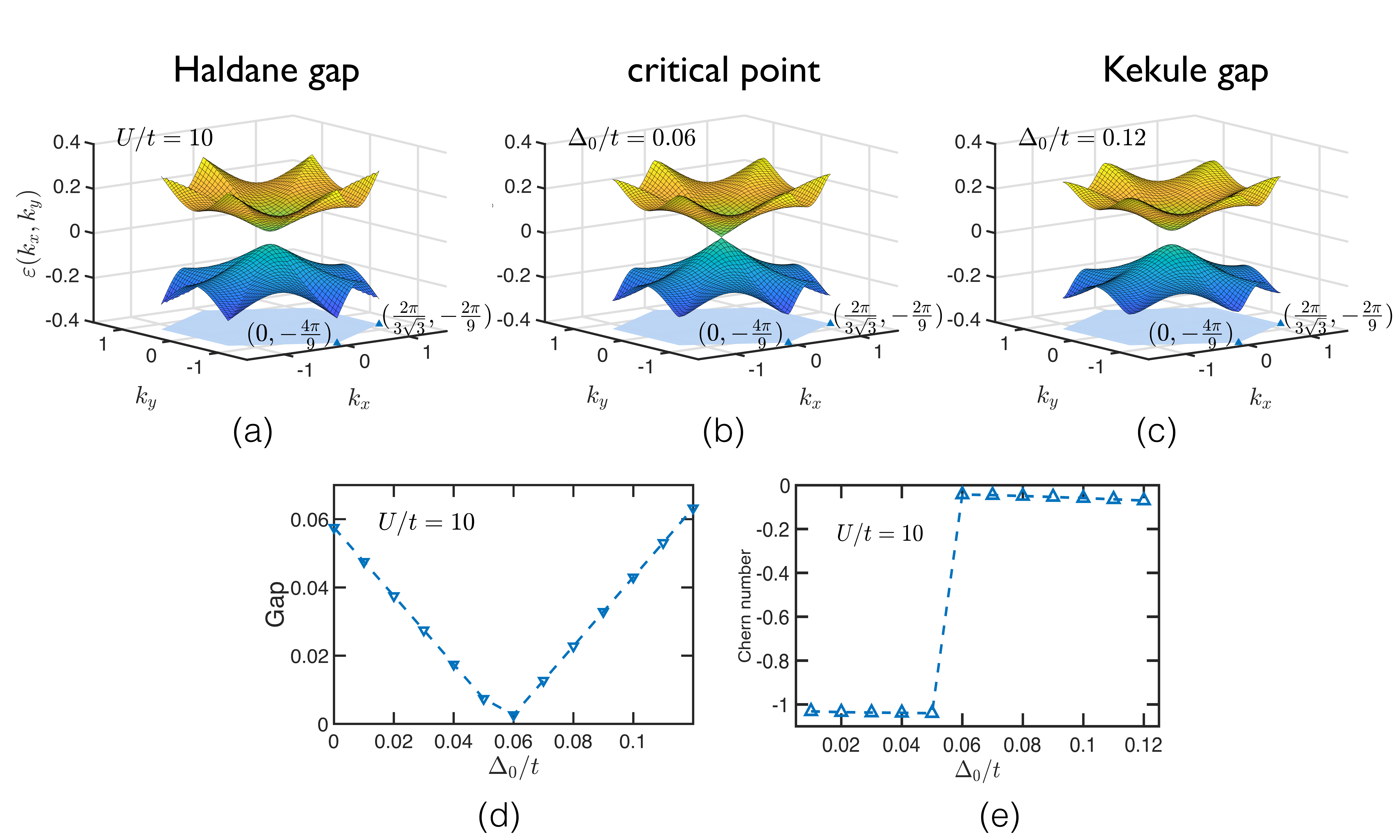}
\caption{
Upper panels:
the pair of particle-hole symmetric bands with the lowest energies for
Hamiltonian~(\ref{eq:wire}) when ${{U}}/{{t}}=10$ with ${{U}}>{{t}}>0$
in the reduced Brillouin zone
$\Omega^{\mathrm{K}}_{\mathrm{BZ}}$ (depicted in light blue).
The energies are plotted in units of $t$.
(a) Haldane gap at the corners of the original Brillouin zone
$\Omega^{\,}_{\mathrm{BZ}}$ in the absence of Kekul\'e dimerization is folded to
the $\Gamma$ point of $\Omega^{\mathrm{K}}_{\mathrm{BZ}}$.
(b) The critical point where the gap closes when
$\Delta^{\,}_{0}/{{t}}\approx 0.06$.
(c) A Kekul\'e gap is present at the $\Gamma$ point in the reduced
Brillouin zone for $\Delta^{\,}_{0}/{{t}}=0.12$.
Lower panel:
(d) the single-particle spectral gap as a function of
$\Delta^{\,}_{0}/{{t}}$. Upon increasing $\Delta^{\,}_{0}/{{t}}$, the gap first
closes and then reopens, indicating a phase transition separating two
distinct gapped phases in which either the Haldane gap or the Kekul\'e
gap dominates; (e) Chern number of the bottom band as a function of $\Delta_0/t$. Across the phase transition, the Chern number jumps from $C=-1$ to $C=0$.
        }
\label{fig:band haldane} 
\end{figure*}

When the system is perturbed by a Kekul\'e dimerization defined by
\begin{subequations}
\label{eq: first def kekule dimerization}
\begin{equation}
\delta\widehat{H}^{\,}_{\mathrm{dimer}}\:= \mathrm{i}
\sum_{j\in\Lambda^{\,}_{A}}
\sum_{\alpha=\mathtt{x},\mathtt{y},\mathtt{z}}
\delta {{t}}_{j, \alpha}\,
\hat{\gamma}^{\alpha}_{A,j}\,
\hat{\gamma}^{\alpha}_{B,j+\bm{s}^{\,}_{\alpha}}
\label{eq: first def kekule dimerization a}  
\end{equation}
with the dimerization pattern%
~\cite{chamon}
\begin{equation}
\delta{{t}}^{\,}_{j,\alpha}\:=
\Delta\,
e^{\mathrm{i}\bm{K}^{\,}_{+}\cdot\bm{s}^{\,}_{\alpha}}\,
e^{\mathrm{i}\bm{G}\cdot\bm{r}_j}
+
\mathrm{c.c.},
\label{eq: first def kekule dimerization b} 
\end{equation}
where the Kekul\'e amplitude
\begin{equation} 
\Delta\:=\Delta^{\,}_{0}\, e^{\mathrm{i}\varphi},
\qquad
\Delta^{\,}_{0}\:=|\Delta|,
\qquad
\varphi\in[0,2\pi),
\label{eq: first def kekule dimerization c} 
\end{equation}
and
\begin{equation}
\bm{G}\:=
\bm{K}^{\,}_{+}-\bm{K}^{\,}_{-}\equiv
2\bm{K}^{\,}_{+}\equiv
-2\bm{K}^{\,}_{-}
\label{eq: first def kekule dimerization d} 
\end{equation}
\end{subequations}
is the momentum connecting the two valleys,
such that
$0<\Delta^{\,}_{0}\ll{{t}}^{2}/{{U}}$,
the band gap decreases until it
vanishes when $\Delta^{\,}_{0}\sim{{t}}^{2}/{{U}}$.
When the Kekul\'e amplitude
$\Delta^{\,}_{0}\gtrsim {{t}}^{2}/{{U}}$,
the gap is of Kekul\'e character~\cite{hou2007}.
This case is illustrated in Fig.\  \ref{fig:band haldane}.
We stress that the Haldane and Kekul\'e gaps compete against each other,
so they realize two distinct gapped phases separated by a gap-closing
transition.~\cite{shinsei}
When the gap is of Haldane character, the bottom band has a Chern number $C=-1$ and there is a chiral mode propagating along the edge of a system with boundary. On the other hand, the Chern number vanishes across the phase transition when the gap is dominated by Kekul\'e dimerization, as shown in Fig.~\ref{fig:band haldane}(e). In this phase, there is no chiral edge mode at the boundary of the system~\cite{PhysRevLett.100.110405}.

\subsection{Scaling limits}
\label{subsec: Scaling limits}

There is an interesting scaling limit of
(\ref{eq: def H honeycomb nanonwires})
consisting in taking the limit ${{U}}\to\infty$ holding ${{t}}$ fixed.
In this limit, the hierarchy
\begin{subequations}
\label{eq: hierarchy}
\begin{equation}
{{U}}>{{t}}>\frac{{{t}}}{{{U}}}\,{{t}}
\label{eq: hierarchy a}
\end{equation}
becomes
\begin{equation}
  \infty>{{t}}>0.
\label{eq: hierarchy b}
\end{equation}
\end{subequations}
This limit sends to infinite energy the two pairs of
particle-hole symmetric Majorana bands that are separated by
an energy of order $2{{U}}$ [see Eqs.~\eqref{eq:trimer solution}].
It leaves a \textit{gapless}
pair of particle-hole symmetric Majorana bands with conical
band crossing at the corners $\bm{K}^{\,}_{+}$ and $\bm{K}^{\,}_{-}$
of the Brillouin zone $\Omega^{\,}_{\mathrm{BZ}}$. In this limit,
time-reversal symmetry, as measured by the vanishing of the Haldane
gap, is restored. This limit is useful as it allows one to treat
in closed analytical form the effect of a Kekul\'e modulation of
${{t}}$ -- in particular the effect of a vortex in the Kekul\'e modulation of
${{t}}$ -- on the single-particle spectrum. 

\section{Free Majoranas on a honeycomb lattice with Kekul\'e dimerization}
\label{sec: free majorana}

We start by reviewing the properties of a tight-binding model for
Majoranas hopping on the honeycomb lattice with nearest-neighbor hopping
amplitudes. This model is motivated by the
scaling limit ${{U}}\to\infty$ holding ${{t}}$ fixed that turns the hierarchy
(\ref{eq: hierarchy a})
into the hierarchy
(\ref{eq: hierarchy b}).

\subsection{Gapless liquid phase with uniform hopping amplitudes}

Consider a honeycomb lattice $\Lambda$
made of two interpenetrating triangular lattices
$\Lambda^{\,}_{A}$
and
$\Lambda^{\,}_{B}$.
We start with the operator
$\hat{a}^{\,}_{\bm{r}}$
that either creates or annihilates a Majorana mode
on the lattice site $\bm{r}$, i.e.,
\begin{subequations}
\begin{equation}
\{\hat{a}^{\,}_{\bm{r}},\hat{a}^{\,}_{\bm{r}'}\}=
2\delta^{\,}_{\bm{r},\bm{r}'},
\qquad
\hat{a}^{\dag}_{\bm{r}}=\hat{a}^{\,}_{\bm{r}},
\end{equation}
for any pair of sites $\bm{r}$ and $\bm{r}'$.
We endow these Majorana modes with the quantum dynamics
specified by the single-particle Hamiltonian
\begin{equation}
\widehat{H}\:=
\sum_{\bm{r}\in\Lambda^{\,}_{A}}
\sum_{\alpha=\mathtt{x},\mathtt{y},\mathtt{z}}
{{t}}\,
\mathrm{i}
\hat{a}^{\,}_{\bm{r}}\,
\hat{a}^{\,}_{\bm{r}+\bm{s}^{\,}_{\alpha}}.
\label{eq:free maj}
\end{equation}
\end{subequations}
Without loss of generality, we choose the hopping amplitudes to be positive,
${{t}}>0$.
We have set the lattice spacing $\mathfrak{a}$ of the honeycomb
lattice to unity, $\mathfrak{a}=1$.

We observe that Majorana operators localized on sublattice $\Lambda^{\,}_{A}$
always appear to the left of Majorana operators localized on sublattice
$\Lambda^{\,}_{B}$ in the Hamiltonian (\ref{eq:free maj}).
If we define the operation of time reversal by the rule
\begin{equation}
\mathrm{i}\mapsto-\mathrm{i},
\qquad
\hat{a}^{\,}_{\bm{r}}\mapsto+\hat{a}^{\,}_{\bm{r}},
\qquad
\hat{a}^{\,}_{\bm{r}+\bm{s}^{\,}_{\alpha}}\mapsto-\hat{a}^{\,}_{\bm{r}+\bm{s}^{\,}_{\alpha}},
\label{eq: def reversal time for a's}
\end{equation}
we conclude that the Hamiltonian (\ref{eq:free maj}) is invariant
under reversal of time. 

Hamiltonian~(\ref{eq:free maj})
is invariant under the translations that map the honeycomb lattice
onto itself. Hence, we perform the Fourier transformation
\begin{subequations}
\label{eq: def Fourier trsf Majorana}
\begin{align}
&
\hat{a}^{\,}_{\bm{r}}\=:
\frac{1}{\sqrt{N}}
\sum_{\bm{k}\in\Omega^{\,}_{\mathrm{BZ}}}
e^{\mathrm{i}\bm{k}\cdot\bm{r}}
\hat{a}^{\,}_{A,\bm{k}},
\\
&
\hat{a}^{\,}_{\bm{r}+\bm{s}^{\,}_{\alpha}}\=:
\frac{1}{\sqrt{N}}
\sum_{\bm{k}\in\Omega^{\,}_{\mathrm{BZ}}}
e^{\mathrm{i}\bm{k}\cdot(\bm{r}+\bm{s}^{\,}_{\alpha})}
\hat{a}^{\,}_{B,\bm{k}},
\end{align}
where $\Omega^{\,}_{\mathrm{BZ}}$ denotes the Brillouin zone of
the triangular sublattice. Notice that since
$\hat{a}^{\,}_{\bm{r}}$ is a Majorana operator,
$\hat{a}^{\dag}_{\bm{k}}$ and $\hat{a}^{\,}_{\bm{k}}$ are not
independent,
\begin{equation}
\hat{a}^{\dag}_{A,\bm{k}}=\hat{a}^{\,}_{A,-\bm{k}},
\qquad
\hat{a}^{\dag}_{B,\bm{k}}=\hat{a}^{\,}_{B,-\bm{k}}.
\label{eq: Majorana reality in momentum space} 
\end{equation}
\end{subequations}
If we introduce the two-component spinor
\begin{subequations}
\begin{equation}
\hat{\gamma}^{\dag}_{\bm{k}}\:=
\begin{pmatrix}
\hat{a}^{\dag}_{A,\bm{k}} & \hat{a}^{\dag}_{B,\bm{k}}
\end{pmatrix},
\label{eq: honeycomb dispersion a}
\end{equation}
Hamiltonian~(\ref{eq:free maj}) turns into
\begin{equation}
\widehat{H}=
\sum_{\bm{k}\in\Omega^{\,}_{\mathrm{BZ}}}
\hat{\gamma}^{\dag}_{\bm{k}}\,
\mathrm{i}\mathcal{A}^{\,}_{\bm{k}}\,
\hat{\gamma}^{\,}_{\bm{k}},
\label{eq: honeycomb dispersion b}
\end{equation}
where
\begin{equation}
\begin{split}
\mathcal{H}^{\,}_{\bm{k}}\equiv&\,
\mathrm{i}\mathcal{A}^{\,}_{\bm{k}}
\\
\:=&\,
\frac{\mathrm{i}}{2}
\begin{pmatrix}
0
&
+
{{t}}
\sum\limits_{\alpha=\mathtt{x},\mathtt{y},\mathtt{z}}
e^{-\mathrm{i}\bm{k}\cdot\bm{s}^{\,}_{\alpha}}
\\
-
{{t}}
\sum\limits_{\alpha=\mathtt{x},\mathtt{y},\mathtt{z}}
e^{+\mathrm{i}\bm{k}\cdot\bm{s}^{\,}_{\alpha}}
&
0
\end{pmatrix}.
\end{split}
\label{eq: honeycomb dispersion c}
\end{equation}
\end{subequations}

We observe that the symmetry under reversal of time defined by 
Eq.\ (\ref{eq: def reversal time for a's})
is broken by adding to the single-particle Hamiltonian
(\ref{eq: honeycomb dispersion c}) 
the traceless diagonal matrix
\begin{subequations}
\label{eq: def Haldane term}
\begin{equation}
\mathcal{H}^{\mathrm{Hal}}_{\bm{k}}\:=
\begin{pmatrix}
+\Delta^{\mathrm{Hal}}_{\bm{k}}
&
0
\\
0
&
-\Delta^{\mathrm{Hal}}_{\bm{k}}
\end{pmatrix},
\end{equation}
where we demand that the so-called Haldane amplitude satisfies
\begin{equation}
\Delta^{\mathrm{Hal}}_{-\bm{k}}=
-\Delta^{\mathrm{Hal}}_{+\bm{k}}
\end{equation}
\end{subequations}
for any $\bm{k}$ in the Brillouin zone $\Omega^{\,}_{\mathrm{BZ}}$.

Solving for
\begin{subequations}
\begin{equation}
\sum\limits_{\alpha=\mathtt{x},\mathtt{y},\mathtt{z}}
e^{+\mathrm{i}\bm{k}\cdot\bm{s}^{\,}_{\alpha}}=0
\end{equation}
yields the two nodal points 
\begin{equation}
\bm{K}^{\,}_{\pm}\:=
\frac{4\pi}{3\sqrt{3}}
\begin{pmatrix}
\pm 1
\\
0
\end{pmatrix}
\label{eq:kpm}
\end{equation}
\end{subequations}
at the corners $\bm{K}^{\,}_{\pm}$ of the Brillouin zone.
Hence, the single-particle spectrum
of the single-particle Hamiltonian
(\ref{eq: honeycomb dispersion c})
is identical to that of graphene for spinless fermions
at vanishing chemical potential by virtue of the Majorana
representation in the second-quantized Hamiltonian
(\ref{eq: honeycomb dispersion b}).

Adding the Haldane term
(\ref{eq: def Haldane term})
to the single-particle Hamiltonian
(\ref{eq: honeycomb dispersion c})
opens a gap $2|\Delta^{\,}_{\bm{K}^{\,}_{+}}|>0$, the so-called Haldane gap,
at the corners $\bm{K}^{\,}_{\pm}$ of the Brillouin zone.
The upper and lower bands carry opposite Chern numbers of magnitude $1$
when $2|\Delta^{\,}_{\bm{K}^{\,}_{+}}|>0$. This Haldane gap is the counterpart to
the gap (\ref{eq: Haldane gap nanowire network}).

If we focus on the low-energy physics near the two Majorana cones,
we can write
$\bm{k}=\bm{K}^{\,}_{\pm}+\bm{p}$
in the vicinity of
$\bm{K}^{\,}_{\pm}$
and expand to leading order in
$\bm{p}$.
The linearized Hamiltonian~(\ref{eq:free maj}) now takes the form
\begin{subequations}
\begin{align}
&
\widehat{H}\approx
\frac{1}{2}
\int\limits_{\Omega^{\,}_{\mathrm{BZ}}}\frac{\mathrm{d}^{2}\bm{p}}{(2\pi)^{2}}\,
\widehat{\Upsilon}^{\dag}(\bm{p})\,
\mathrm{i}\widetilde{\mathcal{A}}(\bm{p})\,
\widehat{\Upsilon}(\bm{p}),
\\
&
\widetilde{\mathcal{H}}(\bm{p})\equiv
\mathrm{i}\widetilde{\mathcal{A}}(\bm{p})=
v^{\,}_{\mathrm{F}}\,
\begin{pmatrix}
-\bm{p}\cdot\bm{\sigma}&0
\\
0&+\bm{p}\cdot\bm{\sigma}
\end{pmatrix},
\end{align}
where $v^{\,}_{\mathrm{F}}\:=3{{t}}/2$ and $\bm{\sigma}$ are Pauli matrices
acting on the two sublattice degrees of freedom. We have introduced
the four-component spinor
\begin{equation} 
\widehat{\Upsilon}^{\dag}(\bm{p}) =
\begin{pmatrix}
\hat{a}^{\dag}_{A,+}(\bm{p})
&
-\mathrm{i}\hat{a}^{\dag}_{B,+}(\bm{p})
&
-\mathrm{i}\hat{a}^{\dag}_{B,-}(\bm{p})
&
\hat{a}^{\dag}_{A,-}(\bm{p})
\end{pmatrix},
\end{equation}
where the subscript $\pm$ labels the two valleys centered about
the nodal points (\ref{eq:kpm}).
If we introduce another set of Pauli matrices $\bm{\tau}$ acting on
these valley degrees of freedom, the constraint from the reality condition
becomes
\begin{equation} 
\widehat{\Upsilon}^{\dag}(\bm{p})=
[-\sigma^{2}\otimes\tau^{2}\widehat{\Upsilon}(-\bm{p})]^{\mathsf{T}}.
\label{eq:particle hole}
\end{equation}
\end{subequations}

If we do the rescaling
\begin{subequations}
\label{eq:free dirac}
\begin{equation}
\widehat{\Upsilon}^{\dag}(\bm{p})\=:
\sqrt{2}\,\widehat{\Psi}^{\dag}(\bm{p}),
\qquad
\widehat{\Upsilon}^{\,}(\bm{p})\=:
\sqrt{2}\,\widehat{\Psi}^{\,}(\bm{p}),
\label{eq:free dirac a}
\end{equation}
one may verify that the components of $\widehat{\Psi}^{\dag}(\bm{p})$
obey the standard algebra of \textit{complex} fermions in momentum space
within each valley subspace. Finally, we arrive at the representation
\begin{align}
&
\widehat{H}\approx
\int
\frac{\mathrm{d}^{2}\bm{p}}{(2\pi)^{2}}\,
\widehat{\Psi}^{\dag}(\bm{p})\,
\widetilde{\mathcal{H}}(\bm{p})
\widehat{\Psi}(\bm{p}),
\label{eq:free dirac b}
\\
&
\widetilde{\mathcal{H}}(\bm{p})\equiv
-
v^{\,}_{\mathrm{F}}\,
\bm{p}\cdot\bm{\sigma}\otimes\tau^{3},
\label{eq:free dirac c}
\\
&
\widehat{\Psi}^{\dag}(\bm{p})=
[-\sigma^{2}\otimes\tau^{2}\widehat{\Psi}(-\bm{p})]^{\mathsf{T}}.
\label{eq:free dirac d}
\end{align}
\end{subequations}
This is the same Hamiltonian as the one governing the vortex-free sector
of Kitaev's honeycomb model.%
~\cite{kitaev2006}
The spinors $\widehat{\Psi}(\bm{p})$ and $\widehat{\Psi}^{\dag}(\bm{p})$
are not independent due to the constraint (\ref{eq:free dirac d}),
which is essentially a particle-hole constraint that relates the
operators at one valley to the other valley. Therefore, the single-particle
Hamiltonian (\ref{eq:free dirac c}) has a BdG form.

\begin{figure*}[t]
\centering
(a)
\includegraphics[width=.45\textwidth]{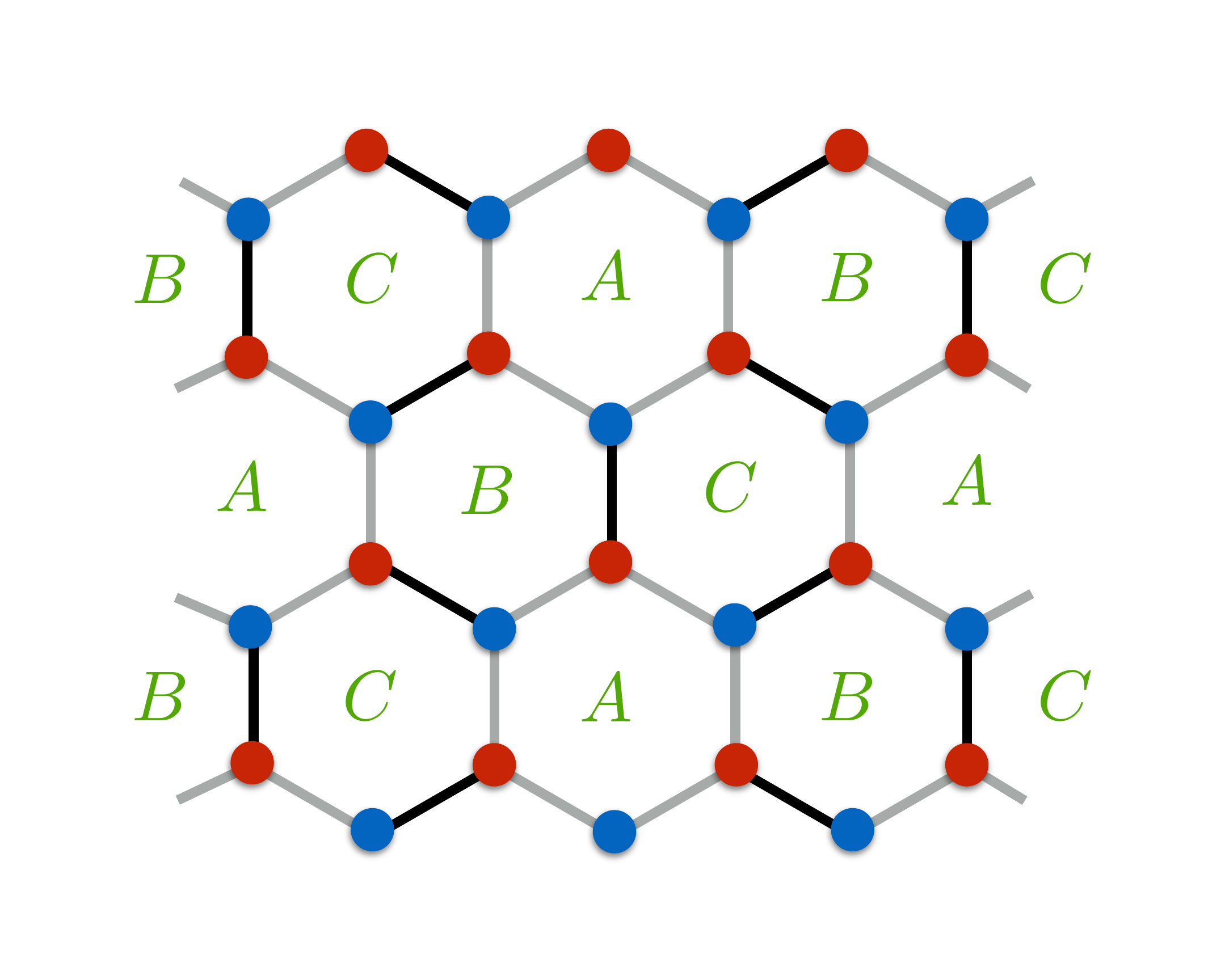}
(b)
\includegraphics[width=.34\textwidth]{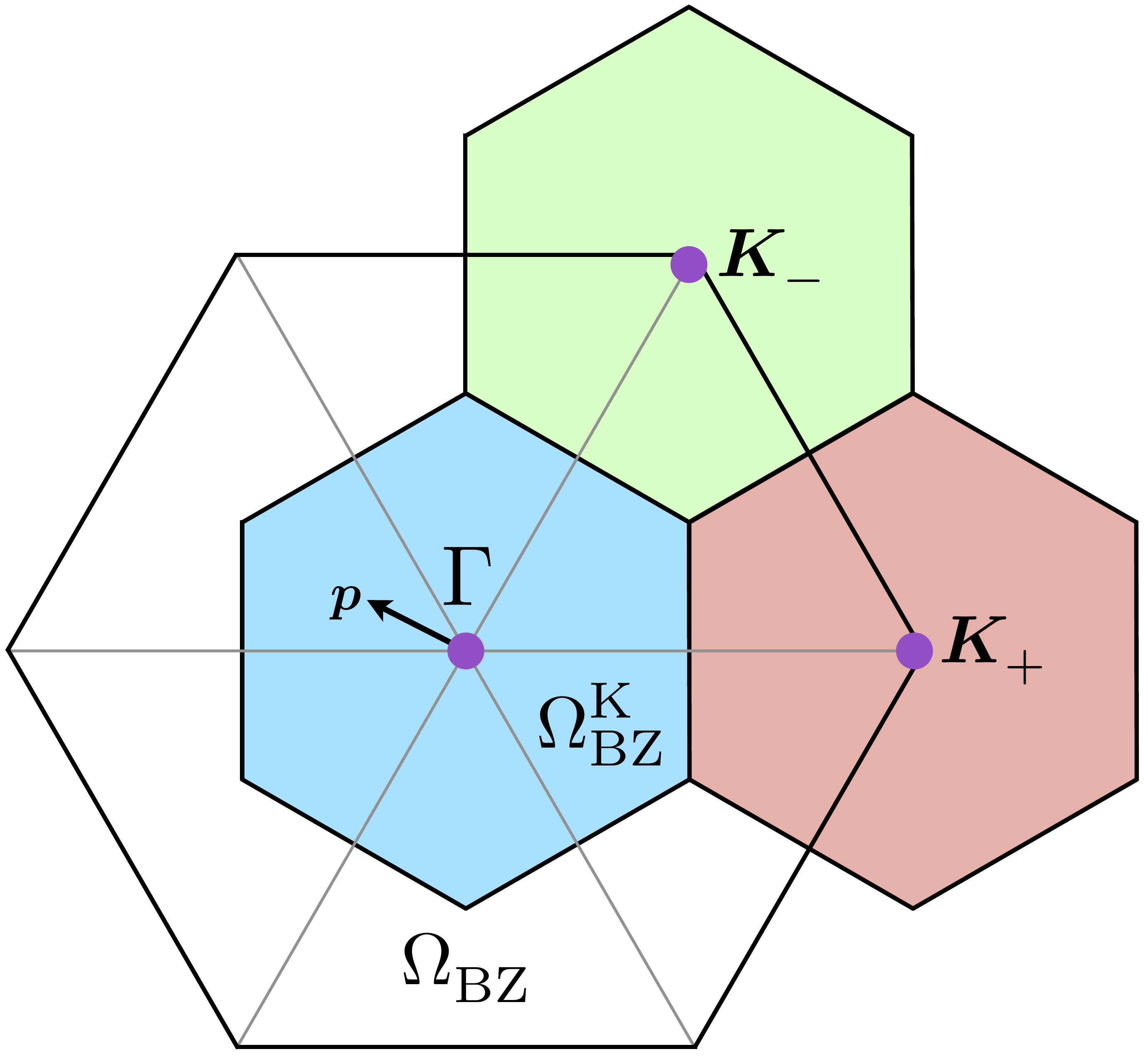}
\caption{
(a) 
The Kekul\'e modulation of the coupling strengths along the bonds. The
black (grey) color denotes hopping amplitudes
that are strong (weak). Such a dimerization pattern breaks the space
group symmetry of the original Bravais lattice by enlarging the original
unit cell. We label the inequivalent plaquettes by $A$, $B$, and $C$, and
the enlarged unit cell is made of three original unit cells.
(b)
Folding the Brillouin zone $\Omega^{\,}_{\mathrm{BZ}}$ of the
honeycomb lattice into the Kekul\'e Brillouin zone
$\Omega^{\mathrm{K}}_{\mathrm{BZ}}$.  The three colored Brillouin
zones are equivalent up to translation by reciprocal lattice vectors
of the folded Brillouin zone.
         }
\label{fig:kekule} 
\end{figure*}

\subsection{Gapped phase with Kekul\'e dimerization}

We consider the effect of a Kekul\'e modulation of the
hopping amplitudes along the bonds of the
honeycomb lattice. As we will see, the Kekul\'e dimerization will open
a gap near $\bm{K}^{\,}_{\pm}$. The Hamiltonian describing the Kekul\'e
modulation can be represented by
[compare with Eq.\ (\ref{eq: first def kekule dimerization})]
\begin{subequations}
\label{eq: def Kekule perturbation}
\begin{equation}
\delta\widehat{H}\:=
\mathrm{i}
\sum_{{\bm r}\in\Lambda^{\,}_{A}}
\sum_{\alpha=\mathtt{x},\mathtt{y},\mathtt{z}}
\delta{{t}}^{\,}_{\bm{r},\alpha}\,
\hat{a}^{\,}_{\bm{r}}\,
\hat{a}^{\,}_{\bm{r}+\bm{s}^{\,}_{\alpha}}
\end{equation}
with the dimerization pattern%
~\cite{chamon}
\begin{equation}
\delta{{t}}^{\,}_{\bm{r},\alpha}\:=
\frac{\Delta}{3}\,
e^{\mathrm{i}\bm{K}^{\,}_{+}\cdot\bm{s}^{\,}_{\alpha}}\,
e^{\mathrm{i}\bm{G}\cdot\bm{r}}
+
\mathrm{c.c.},
\label{eq:kekule}
\end{equation}
where the Kekul\'e amplitude
\begin{equation} 
\Delta\:=\Delta^{\,}_{0}\, e^{\mathrm{i}\varphi},
\qquad
\Delta^{\,}_{0}\:=|\Delta|,
\qquad
\varphi\in[0,2\pi),
\end{equation}
\end{subequations}
will be shown to be associated to a single-particle gap
that opens up at the nodal points (\ref{eq:kpm}). 
The Kekul\'e term (\ref{eq:kekule})
modulates the magnitudes of the hopping amplitudes
along the bonds in an
alternating fashion as shown in Fig.~\ref{fig:kekule}(a). Such a
dimerization pattern breaks the space group symmetry of the original
Bravais lattice by enlarging the original unit cell. In Fig.\
\ref{fig:kekule}(a),
we label the inequivalent plaquettes by $A$, $B$, and $C$. By inspection
of Fig.\ \ref{fig:kekule}(a), one observes
that the enlarged unit cell is made of three original ones.
As a result, we now have a smaller Brillouin zone
$\Omega^{\mathrm{K}}_{\mathrm{BZ}}$
corresponding to the enlarged unit cell, see Fig.\ \ref{fig:kekule}(b).
There are $3\times2=6$ Majorana bands with
all momenta from the original Brillouin zone $\Omega^{\,}_{\mathrm{BZ}}$
folded into $\Omega^{\mathrm{K}}_{\mathrm{BZ}}$. Applying the Fourier
transformation~(\ref{eq: def Fourier trsf Majorana}), the Kekul\'e
modulation~(\ref{eq: def Kekule perturbation}) takes the form
\begin{widetext}
\begin{equation}
\delta\widehat{H}=\mathrm{i}
\sum_{\bm{k}\in\Omega^{\,}_{\mathrm{BZ}}}
\left[
\left(
\sum_{\alpha=\mathtt{x},\mathtt{y},\mathtt{z}}
\frac{2\Delta}{3}\,
e^{\mathrm{i}(\bm{K}^{\,}_{+}+\bm{k})\cdot\bm{s}^{\,}_{\alpha}}
\right)
\hat{a}^{\dag}_{A,[\bm{k}+\bm{G}]}\,
\hat{a}^{\,}_{B,[\bm{k}]}
+
\left(
\sum_{\alpha=\mathtt{x},\mathtt{y},\mathtt{z}}
\frac{2\overline{\Delta}}{3}\,
e^{\mathrm{i}(\bm{K}^{\,}_{-}+\bm{k})\cdot\bm{s}^{\,}_{\alpha}}
\right)
\hat{a}^{\dag}_{A,[\bm{k}-\bm{G}]}\,
\hat{a}^{\,}_{B,[\bm{k}]}
\right],
\label{eq:kekule fourier}
\end{equation}
where we have used the reality condition
(\ref{eq: Majorana reality in momentum space})
and defined $[\bm{q}]$ as the wave vector in
the union of the three colored hexagonal cells in
Fig.~\ref{fig:kekule}(b)
that differs from $\bm{q}$ by a reciprocal wave vector.
Expanding Eq.~(\ref{eq:kekule fourier})
near $\bm{K}_{^{\,}_{\pm}}$ and the $\Gamma$ point, we obtain
\begin{align}
\delta\widehat{H}=&\,
\mathrm{i}
\left[
\sum_{\bm{p}\in \Omega^{\mathrm{K}}_{\mathrm{BZ}}}
\left(
\sum_{\alpha=\mathtt{x},\mathtt{y},\mathtt{z}}
\frac{2\Delta}{3}\,
e^{\mathrm{i}(2\bm{K}^{\,}_{+}+\bm{p})\cdot\bm{s}^{\,}_{\alpha}}
\right)
\hat{a}^{\dag}_{A,[3\bm{K}^{\,}_{+}+\bm{p}]}\,
\hat{a}^{\,}_{B,[\bm{K}^{\,}_{+}+\bm{p}]}
+
\sum_{\bm{p}\in \Omega^{\mathrm{K}}_{\mathrm{BZ}}}
\left(
\sum_{\alpha=\mathtt{x},\mathtt{y},\mathtt{z}}
\frac{2\overline{\Delta}}{3}\,
e^{\mathrm{i}\bm{p}\cdot\bm{s}^{\,}_{\alpha}}
\right)
\hat{a}^{\dag}_{A,[\bm{K}^{\,}_{-}+\bm{p}]}\,
\hat{a}^{\,}_{B,[\bm{K}^{\,}_{+} +\bm{p}]}
\right.
\nonumber\\
&\,
+
\left.
\sum_{\bm{p}\in \Omega^{\mathrm{K}}_{\mathrm{BZ}}}
\left(
\sum_{\alpha=\mathtt{x},\mathtt{y},\mathtt{z}}
\frac{2\Delta}{3}\,
e^{\mathrm{i}\bm{p}\cdot\bm{s}^{\,}_{\alpha}}
\right)
\hat{a}^{\dag}_{A,[\bm{K}^{\,}_{+}+\bm{p}]}\,
\hat{a}^{\,}_{B,[\bm{K}^{\,}_{-}+\bm{p}]}
+
\sum_{\bm{p}\in \Omega^{\mathrm{K}}_{\mathrm{BZ}}}
\left(\sum_{\alpha=\mathtt{x},\mathtt{y},\mathtt{z}}
\frac{2\overline{\Delta}}{3}\,
e^{\mathrm{i}(2\bm{K}^{\,}_{-}+\bm{p})\cdot\bm{s}^{\,}_{\alpha}}
\right)
\hat{a}^{\dag}_{A, [3\bm{K}^{\,}_{-}+\bm{p}]}\,
\hat{a}^{\,}_{B, [\bm{K}^{\,}_{-} +\bm{p}]}
\right.
\nonumber\\
&\,
+
\left.
\sum_{\bm{p}\in \Omega^{\mathrm{K}}_{\mathrm{BZ}}}
\left(
\sum_{\alpha=\mathtt{x},\mathtt{y},\mathtt{z}}
\frac{2\Delta}{3}\,
e^{\mathrm{i}(\bm{K}^{\,}_{+}+\bm{p})\cdot\bm{s}^{\,}_{\alpha}}
\right)
\hat{a}^{\dag}_{A,[\bm{K}^{\,}_{-}+\bm{p}]}\,
\hat{a}^{\,}_{B, [\bm{p}]}
+
\sum_{\bm{p}\in \Omega^{\mathrm{K}}_{\mathrm{BZ}}}
\left(\sum_{\alpha=\mathtt{x},\mathtt{y},\mathtt{z}}
\frac{2\overline{\Delta}}{3}\,
e^{\mathrm{i}(\bm{K}^{\,}_{-}+\bm{p})\cdot\bm{s}^{\,}_{\alpha}}
\right)
\hat{a}^{\dag}_{A, [\bm{K}^{\,}_{+}+\bm{p}]}\,
\hat{a}^{\,}_{B, [\bm{p}]}
\right].
\label{eq:six band}  
\end{align}
The modes at the $\Gamma$ point must also be taken into account, since
the expansion near $\bm{K}^{\,}_{\pm}$ already involves
$3\bm{K}^{\,}_{\pm}$, which can be identified as the $\Gamma$
point.
However, the hybridization
between the nodal modes and the modes at the $\Gamma$ point occurs
at much higher energies.
In the low energy physics, we may neglect terms involving
the modes at the $\Gamma$ point in Eq.~(\ref{eq:six band})
and keep only hybridized modes
between the nodal points $[\bm{K}^{\,}_{\pm}]$. To leading order in
$\bm{p}$, we thus obtain
\begin{subequations}
\begin{equation}
\delta\widehat{H}\approx \mathrm{i}
\int
\frac{\mathrm{d}^{2}\bm{p}}{(2\pi)^{2}}
\left\{
\left[
\Delta\,
\hat{a}^{\dag}_{A,+}(\bm{p})\,
\hat{a}^{\,}_{B,-}(\bm{p})
+
\overline{\Delta}\,
\hat{a}^{\dag}_{A,-}(\bm{p})\,
\hat{a}^{\,}_{B,+}(\bm{p})
\right]
-
\left[
\Delta\,
\hat{a}^{\dag}_{B,+}(\bm{p})\,
\hat{a}^{\,}_{A,-}(\bm{p})
+
\overline{\Delta}\,
\hat{a}^{\dag}_{B,-}(\bm{p})\,
\hat{a}^{\,}_{A,+}(\bm{p})
\right]
\right\},
\end{equation}
where we have made the identifications
\begin{equation}
\hat{a}^{\dag}_{S,[\bm{K}^{\,}_{+}+\bm{p}]}
\to
\hat{a}^{\dag}_{S,+}(\bm{p}),
\qquad
\hat{a}^{\dag}_{S, [\bm{K}^{\,}_{-}+\bm{p}]}
\to
\hat{a}^{\dag}_{S,-}(\bm{p}),
\qquad
S=A,B.
\end{equation}
\end{subequations}
\end{widetext}
Combining with Eq.~(\ref{eq:free dirac b}), the low-energy effective
Hamiltonian in the presence of a Kekul\'e modulation can be written in
the continuum as
\begin{subequations}
\label{eq: def bdg}
\begin{align}
\widehat{H}^{\,}_{\mathrm{Kek}}\:=  
\widehat{H}
+
\delta\widehat{H}
\equiv
\int
\frac{\mathrm{d}^{2}\bm{p}}{(2\pi)^{2}}\,
\widehat{\Psi}^{\dag}(\bm{p})\,
\widetilde{\mathcal{H}}^{\,}_{\mathrm{Kek}}(\bm{p})
\widehat{\Psi}(\bm{p}),
\end{align}
where
\begin{align}
\widetilde{\mathcal{H}}^{\,}_{\mathrm{Kek}}(\bm{p})\:=
\begin{pmatrix}
-\bm{p}\cdot\bm{\sigma}
&
\Delta\,\sigma^{0}
\\
\overline{\Delta}\,\sigma^{0}
&
+\bm{p}\cdot\bm{\sigma}
\end{pmatrix},
\label{eq: def bdg b}
\end{align}
\end{subequations}
and we have set $v^{\,}_{\mathrm{F}}=1$.
We remark that the particle-hole symmetry was never broken on the way
to Eq.~(\ref{eq: def bdg}), so that the single-particle
Hamiltonian (\ref{eq: def bdg b}) is still of the BdG type.
As advertised, the Kekul\'e dimerization opens a gap
$2|\Delta|$ in the single-particle spectrum
due to scattering with the amplitude $\Delta$
between the two nodal points.

\subsection{Symmetry class}

We now consider the symmetries of the BdG Hamiltonian
(\ref{eq: def bdg}). We shall drop the tilde and denote
$\widetilde{\mathcal{H}}(\bm{p})$ simply as $\mathcal{H}(\bm{p})$
from now on.

First, the reality condition (\ref{eq:particle hole}) imposes the
spectral particle-hole symmetry 
\begin{equation}
\mathcal{C}\,
\widetilde{\mathcal{H}}^{\,}_{\mathrm{Kek}}(\bm{p})\,
\mathcal{C}^{-1}=
-\widetilde{\mathcal{H}}^{*}_{\mathrm{Kek}}(-\bm{p}),
\qquad
\mathcal{C}\:=
-\sigma^{2}\otimes\tau^{2}\,\mathsf{K},
\end{equation}
where $\mathsf{K}$ denotes complex conjugation.
Hamiltonian (\ref{eq: def bdg}) also possesses the time-reversal symmetry
\begin{equation}
\mathcal{T}\,
\widetilde{\mathcal{H}}^{\,}_{\mathrm{Kek}}(\bm{p})\,
\mathcal{T}^{-1}=
\widetilde{\mathcal{H}}^{*}_{\mathrm{Kek}}(-\bm{p}),
\qquad
\mathcal{T}\:=\sigma^{1}\otimes\tau^{1}\,\mathsf{K}.
\end{equation}
Finally, composition of $\mathcal{C}$ and $\mathcal{T}$ yields
the chiral symmetry
\begin{equation}
\mathcal{S}\:=
\mathcal{T}\,\mathcal{C}=
\sigma^{3}\otimes\tau^{3},
\label{eq: chiral}
\end{equation}
under which
\begin{equation}
\hat{a}^{\,}_{A}\mapsto\hat{a}^{\,}_{A},
\qquad
\hat{a}^{\,}_{B}\mapsto-\hat{a}^{\,}_{B},
\end{equation}
and
\begin{equation}
\mathcal{S}\,
\widetilde{\mathcal{H}}^{\,}_{\mathrm{Kek}}(\bm{p})\,
\mathcal{S}^{-1}=
-\widetilde{\mathcal{H}}^{\,}_{\mathrm{Kek}}(\bm{p}).
\end{equation}
Notice that the symmetry transformation satisfies
\begin{equation}
\mathcal{C}^{2}=1,
\qquad
\mathcal{T}^{2}=1,
\end{equation}
so that Hamiltonian (\ref{eq: def bdg}) belongs to the symmetry class
BDI. In the presence of point topological defects (vortices), the
Hamiltonian supports zero-energy chiral Majorana modes classified by
$\mathbb{Z}$.%
~\cite{atiyah, jackiw, weinberg, chiu}
As we will see explicitly in the next
section, zero modes with positive and negative chiral eigenvalues have
nonvanishing amplitudes on sublattice $\Lambda^{\,}_{A}$ and $\Lambda^{\,}_{B}$,
respectively.

\subsection{Majorana zero modes bound to Kekul\'e vortices}
\label{subsec: kekule vortex}

The Kekul\'e distortion enters (\ref{eq: def bdg}) as a complex-valued
amplitude. As such the Kekul\'e distortion supports point-like
static defects in the form of vortices
\begin{subequations}
\label{eq: def vtx in Kekule order}
\begin{equation}
\Delta^{\,}_{\mathrm{vtx}}(\bm{r})\:=
\Delta^{\,}_{0}(\bm{r}) e^{\mathrm{i}(\varphi+n\theta)},
\qquad
\bm{r}=
|\bm{r}|
\begin{pmatrix}
\cos\theta
\\
\sin\theta
\end{pmatrix},
\label{eq: def single vortex charge n}
\end{equation}
where $n\in\mathbb{Z}$ is the vorticity that measures the winding of the
phase of the Kekul\'e order parameter, while
$\Delta^{\,}_{0}(\bm{r})\:=|\Delta^{\,}_{\mathrm{vtx}}(\bm{r})|$
defines the profile of its magnitude.
This static function must vanish at
the origin and saturate to some prescribed nonvanishing but finite
value as $\bm{r}\to\infty$, say
\begin{equation}
\Delta^{\,}_{0}(\bm{r})\:=
\Delta^{\,}_{0} 
\tanh\left(\frac{|\bm{r}|}{\ell^{\,}_{0}}\right)
\end{equation}
\end{subequations}
with $\Delta^{\,}_{0}>0$ and $\ell^{\,}_{0}>0$.

We seek any qualitative change induced
in the single-particle spectrum
of Hamiltonian (\ref{eq: def bdg b})
when the Kekul\'e order parameter is given
by Eq.\ (\ref{eq: def vtx in Kekule order})
instead of being a constant complex number.
To this end, we represent Hamiltonian~(\ref{eq: def bdg})
in two-dimensional position space.
We thus have
\begin{subequations}
\label{eq:bdg real space}
\begin{equation}
\widetilde{\mathcal{H}}^{\,}_{\mathrm{Kek}}(\bm{r})\:=
\begin{pmatrix}
  0
  &
  2\mathrm{i}\partial^{\,}_{z}
  &
  \Delta^{\,}_{\mathrm{vtx}}(\bm{r})
  &
  0
  \\
  2\mathrm{i} \partial^{\,}_{\bar{z}}
  &
  0
  &
  0
  & \Delta^{\,}_{\mathrm{vtx}}(\bm{r})
  \\
  \overline{\Delta}^{\,}_{\mathrm{vtx}}(\bm{r})
  &
  0
  &
  0
  &
  -2\mathrm{i}\partial^{\,}_{z}
  \\
  0
  &
  \overline{\Delta}^{\,}_{\mathrm{vtx}}(\bm{r})
  &
  -2\mathrm{i}\partial^{\,}_{\bar{z}}
  &
  0
\end{pmatrix},
\label{eq:bdg real space a}
\end{equation}
where we have chosen the basis
\begin{align}
\widehat{\Psi}^{\dag}(\bm{r})\:=\!\!
\frac{1}{\sqrt{2}}
\begin{pmatrix}
\hat{a}^{\dag}_{A,+}(\bm{r})
&
-\mathrm{i}
\hat{a}^{\dag}_{B,+}(\bm{r})
&
-\mathrm{i}
\hat{a}^{\dag}_{B,-}(\bm{r})
&
\hat{a}^{\dag}_{A,-}(\bm{r})
\end{pmatrix}
\label{eq:bdg real space bb}
\end{align}
obeying the reality condition
\begin{equation}
\widehat{\Psi}^{\dag}(\bm{r})\:= 
\left[-\sigma^{2}\otimes\tau^{2}\widehat{\Psi}(\bm{r})\right]^{\mathsf{T}}
\label{eq:bdg real space b}
\end{equation}
and used the complex coordinates
\begin{equation}
\begin{matrix}
z\:=x+\mathrm{i}y,
\\ \\
\bar{z}\:= x-\mathrm{i}y,
\end{matrix}
\qquad
\begin{matrix}
\partial^{\,}_{z}=
\frac{1}{2}\left(\partial^{\,}_{x}-\mathrm{i}\partial^{\,}_{y}\right),
\\ \\
\partial^{\,}_{\bar{z}}\:=
\frac{1}{2}\left(\partial^{\,}_{x}+\mathrm{i}\partial^{\,}_{y}\right).
\end{matrix}
\label{eq:bdg real space c}
\end{equation}
\end{subequations}
    
We seek normalizable solutions to the eigenvalue problem
\begin{equation}
\mathcal{H}^{\,}_{\mathrm{Kek}}(\bm{r})\,\Psi^{\,}_{0}(\bm{r})=0.
\label{eq:zero mode}
\end{equation}
If a normalizable solution $\Psi^{\,}_{0}(\bm{r})$ exists, we shall call it
a zero mode.
This problem was first studied by Jackiw and Rossi in a different
context where $\Delta(\bm{r})$ is the vortex in the superconducting
order parameter.%
~\cite{jackiw}
Here, the origin of the gap is instead
the bond density wave due to the Kekul\'e modulation.~\cite{hou2007}
Nevertheless,
the mathematical structure of the Hamiltonian%
~(\ref{eq:bdg real space})
is identical to that studied by Jackiw and Rossi. As a
consequence of the spectral chiral symmetry~~(\ref{eq: chiral}), the
single-particle Hamiltonian~(\ref{eq:bdg real space}) is block off
diagonal. Hence, any zero-mode solution must take one of two forms, namely
\begin{align}
&
\Psi^{\,}_{A,0}(\bm{r})=
\begin{pmatrix}
u^{\,}_{A}(\bm{r})
\\
0
\\
0
\\
v^{\,}_{A}(\bm{r}) 
\end{pmatrix},
\qquad
&
\Psi^{\,}_{B,0}(\bm{r})=
\begin{pmatrix}
0
\\
u^{\,}_{B}(\bm{r})
\\
v^{\,}_{B}(\bm{r})
\\
0
\end{pmatrix}.
\end{align}
As is implied by the notation, $\Psi^{\,}_{S,0}(\bm{r})$ has support on
sublattice $S=A, B$ only.
For simplicity, we shall focus below only on cases where $|n|=1$.

When $n=-1$, only $\Psi^{\,}_{A,0}(\bm{r})$ is normalizable.
It is given by
\begin{subequations}
\label{eq: zero mode A}
\begin{align}
&
u^{\,}_{A}(\bm{r})=
\mathcal{N}\,
e^{\mathrm{i}(\frac{\pi}{4}+\frac{\varphi}{2})}\,
e^{-\int\limits_{0}^{r}\mathrm{d}r'\,\Delta^{\,}_{0}(r')},
\\
&
u^{\,}_{B}(\bm{r})=0,
\\
&
v^{\,}_{B}(\bm{r})=0,
\\
&
v^{\,}_{A}(\bm{r})=\overline{u^{\,}_{B}(\bm{r})},
\end{align}
where $\mathcal{N}$ is the normalization constant. The
wavefunction~(\ref{eq: zero mode A})
is exponentially localized about the vortex core,
for it decays exponentially fast with the distance
$\bm{r}$ away from the vortex core with the
characteristic decay length $\sim1/\Delta^{\,}_{0}$ set by the
asymptotic magnitude $\Delta^{\,}_{0}$
of the Kekul\'e order parameter. There follows the ``logical'' MZM operator
\begin{equation}
\hat{\gamma}^{\,}_{A}\:=
\int\mathrm{d}^{2}\bm{r}\,
\left[
u^{\,}_{A}(\bm{r})\,
\hat{a}^{\,}_{A,+}(\bm{r})
+
\overline{u^{\,}_{A}(\bm{r})}\,
\hat{a}^{\,}_{A,-}(\bm{r})
\right].
\label{eq:zero mode operator A}
\end{equation}
The reality condition
\begin{equation}
\gamma^{\dag}_{A}=\gamma^{\,}_{A}
\end{equation}
\end{subequations}
follows from Eq.~(\ref{eq:bdg real space b}).

Similarly, when $n=+1$, it is only $\Psi^{\,}_{B,0}(\bm{r})$
that is normalizable. The wavefunction is then given by
\begin{subequations}
\label{eq: zero mode B}
\begin{align}
&
u^{\,}_{A}(\bm{r})=0,
\\
&
u^{\,}_{B}(\bm{r})=
\mathcal{N}\,
e^{\mathrm{i}(\frac{\pi}{4}+\frac{\varphi}{2})}\,
e^{-\int\limits_{0}^{r}\mathrm{d}r'\,\Delta^{\,}_{0}(r')},
\\
&
v^{\,}_{B}(\bm{r})=\overline{u^{\,}_{A}(\bm{r})},
\\
&
v^{\,}_{A}(\bm{r})=0,
\end{align}
where $\mathcal{N}$ is the normalization constant.
There follows the ``logical'' MZM operator
\begin{equation}
\hat{\gamma}^{\,}_{B}\:=
\int\mathrm{d}^{2}\bm{r}\,
\left[
u^{\,}_{B}(\bm{r})\,
\hat{a}^{\,}_{B,+}(\bm{r})
+
\overline{u^{\,}_{B}(\bm{r})}\,
\hat{a}^{\,}_{B,-}(\bm{r})
\right].
\label{eq:zero mode operator B}
\end{equation}
The reality condition
\begin{equation}
\hat{\gamma}^{\dag}_{B}=\hat{\gamma}^{\,}_{B}
\end{equation}
\end{subequations}
follows from Eq.~(\ref{eq:bdg real space b}).

In summary,
far-separated Kekul\'e vortices with $|n|=1$
bind MZMs localized around their vortex cores,
with nonvanishing amplitude on either
sublattice $\Lambda^{\,}_{A}$ or $\Lambda^{\,}_{B}$, respectively.
For $|n|>1$, the index theorem guarantees that
there are $|n|$ mutually orthogonal normalizable zero modes,
with support on sublattice
$\Lambda^{\,}_{A}$ or $\Lambda^{\,}_{B}$ depending on sgn($n$).%
~\cite{jackiw}
All $n$ zero modes are robust to any perturbation that respects the BDI
symmetry.%
~\cite{atiyah, jackiw, weinberg, chiu}
Thus, in general, Kekul\'{e} vortices in class BDI can harbor
multiple protected MZMs, unlike vortices in the
traditional (2+1)-dimensional $p+\mathrm{i}\,p$
superconductor.%
~\cite{read,ivanov}
The reason for this is that vortices in the latter case carry a
$\mathbb Z^{\,}_{2}$ index,  owing to the fact that the parent Hamiltonian is
in class D rather than BDI, so that only the parity of the number of MZMs
is conserved. The model studied in Sec.\
\ref{Realization with Majorana nanowires}
turns out to be in class D, and consequently is more similar to the usual
$p+\mathrm{i}\, p$ superconductor,
despite the fact that its vortices also stem from the presence of a
Kekul\'{e} distortion.

If we drop the reality condition (\ref{eq:bdg real space b}),
the fermion number becomes a good quantum number. This situtation applies
to the case of complex fermions hopping on the honeycomb lattice
as was considered
in Refs.\ \onlinecite{hou2007,hou2008}. The filled Fermi sea
with the zero mode occupied or empty, respectively, can then be assigned
the fermion number $\pm1/2$. In the presence of the reality
condition (\ref{eq:bdg real space b}), the zero mode becomes a logical MZM
of indefinite fermion number.
The logical MZMs obey an exotic braiding statistics, as we now explain.

\subsection{Braiding statistics of Kekul\'{e} vortices}
\label{subsec: braiding statistics}

In this section, we review the fact that the form of the zero-mode solutions
(\ref{eq:zero mode operator A}) and (\ref{eq:zero mode operator B})
implies that their corresponding MZM operators
obey non-Abelian braiding statistics,
just like the half-vortices of $p+\mathrm{i}p$
topological superconductors.%
~\cite{read,ivanov}

Instead of one vortex, we shall consider
$v$ vortices all sharing the same vorticity centered at the positions 
$\bm{R}^{\,}_{1},\cdots,\bm{R}^{\,}_{v}$
on the two-dimensional Euclidean plane
through the Ansatz
\begin{equation}
\begin{split}
\Delta(\bm{r};\bm{R}^{\,}_{1},\cdots,\bm{R}^{\,}_{v})\:=&\,
\Delta^{\,}_{0}
\prod_{j=1}^{v}
\mathrm{tanh}\left(\frac{|\bm{r}-\bm{R}^{\,}_{j}|}{l^{\,}_{0}}\right)
\\
&\,\times
e^{\mathrm{i}[\varphi^{\,}_{j}-\mathrm{arg}(\bm{r}-\bm{R}^{\,}_{j})]}.
\end{split}
\end{equation}
We assume that the vortices are kept far enough away from each other
that their pairwise hybridization can be ignored, i.e.,
\begin{equation}
|\bm{R}^{\,}_{i}-\bm{R}^{\,}_{j}|\gg1/\Delta^{\,}_{0}
\end{equation}
must always hold for any $1\leq i<j\leq v$.
Suppose that $\bm{R}^{\,}_{j}$ moves adiabatically anticlockwise once along
a closed path in two-dimensional Euclidean space.
Furthermore, suppose that this path encircles one and only one
vortex, say the vortex located at $\bm{R}^{\,}_{i}$ without loss of
generality.  If $\bm{r}$ is sufficiently close to $\bm{R}^{\,}_{i}$,
$\mathrm{arg}(\bm{r}-\bm{R}^{\,}_{j})$ changes by $2\pi$, a change
that can be absorbed by taking
$\varphi^{\,}_{i}\rightarrow\varphi^{\,}_{i}+2\pi$.
However, due to the presence of the phase $\varphi^{\,}_{i}/2$
in the zero mode solutions (\ref{eq: zero mode A}) and
(\ref{eq: zero mode B}),
we find that $\hat{\gamma}^{\,}_{i}\rightarrow-\hat{\gamma}^{\,}_{i}$
after moving $\bm{r}^{\,}_{j}$ a full circle around $\bm{r}^{\,}_{i}$.
Repeating the same analysis by interchanging the role
of $\bm{r}^{\,}_{j}$ and $\bm{r}^{\,}_{i}$, one finds that
$\hat{\gamma}^{\,}_{j}\rightarrow-\hat{\gamma}^{\,}_{j}$ as well.

The appearance of the additional minus sign due to the
multi-valuedness of the zero mode solutions parallels that of the
$p+\mathrm{i}\, p$ topological superconductor. Namely, the MZM operator
changes sign as the vortex phase winds by $2\pi$. To keep track of the
signs, it is convenient to take $\varphi^{\,}_{i}\in[0,2\pi)$ and
introduce branch cuts so that $\varphi^{\,}_{i}$ jumps by $2\pi$ each time
the vortex $\bm{r}^{\,}_{i}$ crosses a branch cut. In this way, one
can derive the following property of the Majorana zero modes under a
counterclockwise exchange of vortices $j$ and $j+1$,
\begin{equation}
\hat{\gamma}^{\,}_{j}\mapsto
+\hat{\gamma}_{j+1},
\qquad
\hat{\gamma}_{j+1}\mapsto
-\hat{\gamma}^{\,}_{j},
\end{equation}
which is precisely the braiding statistics of MZMs.%
~\cite{read,ivanov}

\begin{figure*}[!t]
\centering
\includegraphics[width=0.8\textwidth]{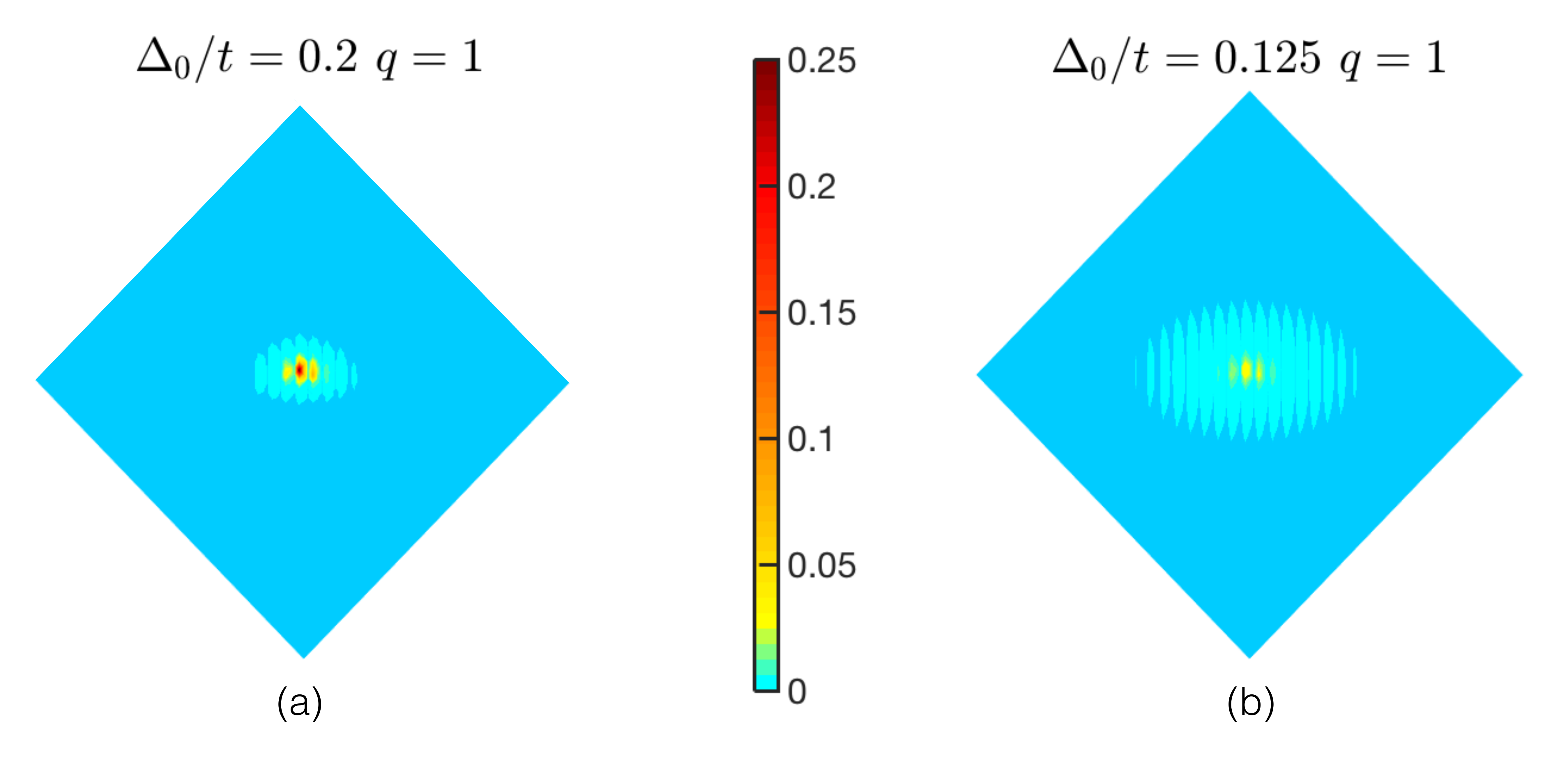}
\caption{
Wavefunctions of the zero mode bound to a single Kekul\'e vortex at
the origin for $\Delta^{\,}_{0}/{{U}}=0.02$ with ${{U}}>{{t}}>0$ and
vorticity $q=1$. The numerics are carried out on a diamond shaped
geometry with 61 sites on each edge. (a) $\Delta_0/t=0.2$; (b)
$\Delta_0/t=0.125$.  The zero mode amplitude decreases upon decreasing
$\Delta_0/t$ and the profile broadens.  For a system with open boundary,
there is an additional zero mode localized near the boundary which is
not shown in the plot.
        }
\label{fig:zero mode haldane} 
\end{figure*}

\section{Zero modes bound to Kekul\'e vortices in the network of Majorana
nanowires}
\label{sec: Zero modes bound to Kekule ...}

We now return to the Hamiltonian (\ref{eq: def H honeycomb nanonwires})
describing the network of quantum nanowires
in the presence of a Kekul\'e gap larger than the Haldane gap.
We shall impose a Kekule vortex of vorticity one in magnitude
and verify numerically that it binds a ``logical'' Majorana zero mode.

To this end, we imprint a Kekul\'e vortex with vorticity
$q=\pm1$ that is centered at the origin, $\bm{R}=\bm{0}$,
by replacing the uniform ${{t}}$ in the dimer Hamiltonian
(\ref{eq:dimer}) with ${{t}}+\delta {{t}}^{\,}_{\bm{r},\alpha}$ where
[compare with Eq.~(\ref{eq:voltage})]
\begin{equation}
\delta {{t}}^{\,}_{\bm{r},\alpha}\:=
\Delta^{\,}_{0}\,
\cos
\left(
\bm{K}^{\,}_{+}\cdot\bm{s}^{\,}_{\alpha}
+
\bm{G}\cdot\bm{r}
+
q\,
\mathrm{arg}\left(\bm{r}\right)
\right)
\label{eq:voltage bis}
\end{equation}
and $\alpha=\mathtt{x},\mathtt{y},\mathtt{z}$.
In the continuum limit, this expression yields a Kekul\'{e} order parameter
with a vortex profile similar to that in
Eq.\ \eqref{eq: def vtx in Kekule order}.

When$|\Delta^{\,}_{0}|\gtrsim {{t}}^{2}/{{U}}$,
we find a zero mode bound to
the Kekul\'e vortex, as shown in Fig.\ \ref{fig:zero mode haldane}.
The amplitude of this zero mode decays exponentially
away from the vortex core.
The amplitudes are nonvanishing on sublattices
$\Lambda^{\,}_{A}$ and $\Lambda^{\,}_{B}$, respectively,
depending on the sign of the vorticity, $\mathrm{sgn}(q)=\pm1$.
Upon increasing ${{t}}/{{U}}$,
the band gap decreases as the
Kekul\'e gap competes with the Haldane gap.
Consequently, the exponential decay of the zero mode is less pronounced,
and the zero mode spreads out further, until the zero mode
is eaten by the continuum of single-particle states when the band gap vanishes.
When ${{t}}^{2}/{{U}}\gtrsim |\Delta^{\,}_{0}|$,
the Haldane gap dominates over the Kekul\'e gap
and no zero mode can bind to a Kekul\'e vortex.%
~\cite{shinsei}

\section{Experimental considerations}
\label{sec: Transporting logical majoranas}

\subsection{Measurement scheme}
\label{subsec: experiment}

We now discuss the possibility of measuring the
emergent MZMs and verifying their braiding properties within the
nanowire network proposed in this paper.
The existence of the ``logical" MZMs can be probed via scanning
tunneling microscopy (STM), where they manifest themselves
as zero-bias peaks in the tunneling differential conductance. In
addition, by employing high-resolution STM conductance mapping
techniques, it is possible to probe the spatial profile of the MZMs,
thereby verifying their localized nature.%
~\cite{nadj1,nadj2,ruby,chevallier}

However, the verification of the existence of the ``logical" MZMs is
not complete unless one can also verify that braiding the ``logical"
MZMs acts on the low-energy Hilbert space of the system in the manner
characteristic of true MZMs.  We now make this idea more precise.  For
a system with $2N$ ``logical" MZMs, each pair of MZMs constitutes a
fermionic state that can be either empty or filled. The fermion
parity (even or odd, respectively) of each pair then specifies the
state of a qubit. Thus, the dimension of the Hilbert space spanned by
the quantum states of these qubits grows as $2^{N-1}$ once the total
fermion parity of the $2N$ MZMs has been fixed.  Braiding ``logical"
MZMs performs unitary transformations on this Hilbert space. Thus, in
order to verify that braiding the ``logical" MZMs acts in the desired
way, one needs a means of measuring the fermion parity of any pair of
MZMs.  Here, we can again exploit the fact that the ``logical" MZMs
can be moved adiabatically by adjusting the array of gate voltages.
Bringing a pair of ``logical" MZMs together by merging two Kekul\'e
vortices effectively ``fuses" the two MZMs. Then, in order to
determine whether the pair of MZMs were in an even- or
odd-fermion-parity state, one can measure the local charge
distribution in the vicinity of the fused pair: if there is a finite
charge density where the two zero modes were fused together, then they
were in an odd-fermion-parity state; if not, then they were in an
even-fermion-parity state. Such a measurement can potentially be
achieved with scanning single-electron transistor microscopy (SSETM),
which can resolve local charge density on the length scale of
nanometers.%
~\cite{yoo,li}
Therefore, in principle, the existence of
MZMs and their braiding and fusion properties can be measured by
interfacing STM and SSETM probes with the nanowire network.

We remark that in practice there are additional practical subtleties when performing measurements with STM or SSETM on our setup. For example, the nanowires used in current experiments are covered by a superconducting shell, which may potentially pose a problem for electron tunneling from STM tips. However, in real experiments the superconducting shell does not cover the entire nanowire, but only on the side of the wire~\cite{lutchyn2018, zhang}. In this way, one can avoid direct contact of the superconducting shell with the STM tip. Similarly, at the Y-junction where three wires meet, one could leave a short segment on each wire uncoated by the superconducting shell. As long as the MZMs at the endpoints have a finite extent, their existence could still potentially be detected by STM/SSETM. In summary, measuring the emergent MZMs experimentally would likely require careful considerations and improvements in experimental techniques but is feasible in principle.

\subsection{Experimental parameters}
\label{subsec: Experimental parameters}

Let $\mathfrak{a}$ be the length of a Majorana nanowire
that we are using as a nearest-neighbor bond of the honeycomb lattice
(i.e., the lattice spacing of the honeycomb network).
We assume that the trimer energy scale ${{U}}$ that enters in
Eq.\ (\ref{eq: def H honeycomb nanonwires})
is
${{U}}\sim\Delta^{\,}_{\mathrm{nw}}$,
so that the physical Majoranas are almost on top of one another.
We seek to express the hopping amplitude ${{t}}$
and the Kekul\'e gap $\Delta^{\,}_{0}$ that enter in
$
\widehat{H}^{\,}_{\mathrm{dimer}}
+\delta\widehat{H}^{\,}_{\mathrm{dimer}}
$
[see Eqs.\
(\ref{eq: def H honeycomb nanonwires})
and
(\ref{eq: first def kekule dimerization})]
in terms of the energy scales entering a single Majorana nanowire.

A single Majorana nanowire wire is modeled as a one-dimensional gas of
non-interacting electrons at the chemical potential $V^{\,}_{\mathrm{g}}$
in proximity to an $s$-wave superconductor, 
whereby the electronic kinetic energy competes with Zeeman,
Rashba spin-orbit, and $s$-wave superconducting pairing
contributions to the Hamiltonian.%
~\cite{lutchyn,oreg,alicea2011}

The expression for the topological gap
$\Delta^{\,}_{\mathrm{nw}}$ of a single Majorana nanowire
is \cite{lutchyn,oreg,alicea2011}
\begin{subequations}
\begin{equation}
\Delta^{\,}_{\mathrm{nw}}\:=
\frac{g\,{{\mu}}^{\,}_{\mathrm{B}}\,|B^{\,}_{z}|}{2}
-
\sqrt{\Delta^{2}_{\mathrm{sc}}+V^{2}_{\mathrm{g}}}>0,
\label{eq: def topological nw gap}
\end{equation}
where $g$ is the effective $g$-factor in the wire,
${{\mu}}^{\,}_{\mathrm{B}}$ is the Bohr magneton,
$|B^{\,}_{z}|$ is the strength of the applied magnetic field along the
Cartesian axis $z$ that is perpendicular to the plane in which the
Majorana nanowires lie,
$\Delta^{\,}_{\mathrm{sc}}$
is the proximity-induced superconducting gap of the Majorana nanowire,
and the gate potential $V^{\,}_{\mathrm{g}}$
sets the chemical potential in the Majorana nanowire.
Physical MZMs are bound to the end points of this Majorana nanowire
if and only if
\begin{equation}
\frac{g{{\mu}}^{\,}_{\mathrm{B}}\,|B^{\,}_{z}|}{2}>
\sqrt{\Delta^{2}_{\mathrm{sc}}+V^{2}_{\mathrm{g}}}.
\end{equation}
\end{subequations}
As the decay length for a physical MZM bound to the end points
of a Majorana nanowire is
\begin{align}
\xi^{\,}_{\mathrm{physical}}=
\frac{\hbar\,v^{\,}_{\mathrm{F},\mathrm{nw}}}{\Delta^{\,}_{\mathrm{nw}}},
\end{align}
where $v^{\,}_{\mathrm{F},\mathrm{nw}}$
is the Fermi velocity of the Majorana nanowire
(which is equal to the spin-orbit coupling in the limit
when the Zeeman energy is much smaller
than the effective electron mass times the spin-orbit coupling
in suitable units),
the overlap between two physical MZMs is then approximately given by
\begin{align}
{{t}}\sim
\frac{
\hbar\,v^{\,}_{\mathrm{F},\mathrm{nw}}
     }
     {
\mathfrak{a}
     }\,
\kappa\,e^{-\kappa},
\qquad
\kappa\:=
\frac{
\mathfrak{a}\,\Delta^{\,}_{\mathrm{nw}}
     }
     {
\hbar\,v^{\,}_{\mathrm{F},\mathrm{nw}}
     },
\label{eq: mu estimate}
\end{align}
when measured in units of energy.
This overlap is controlled by the dimensionless ratio
\begin{subequations}
\label{eq: def kappa}
\begin{equation}
\kappa=
\frac{\mathfrak{a}}{\xi^{\,}_{\mathrm{sc}}}\,
\frac{\Delta^{\,}_{\mathrm{nw}}}{\Delta^{\,}_{\mathrm{sc}}},
\label{eq: def kappa a}
\end{equation}
where we have introduced the proximity-induced superconducting coherence length
\begin{equation}
\xi^{\,}_{\mathrm{sc}}\:=
\frac{\hbar\,v^{\,}_{\mathrm{F},\mathrm{nw}}}{\Delta^{\,}_{\mathrm{sc}}}.
\label{eq: def kappa b}
\end{equation}
\end{subequations}
The overlap ${{t}}$ is thus exponentially suppressed by either
increasing the ratio between the length of the Majorana nanowire
and the proximity-induced superconducting coherence length
or the ratio between the topological gap and the proximity-induced
superconducting gap.

When estimating the size of the Kekul\'e gap $\Delta^{\,}_{0}$,
we assume that we can vary the gate voltages $V^{\,}_{\mathrm{g}}$
along the nearest-neighbor bonds on the honeycomb lattice
by the amount $\delta V^{\,}_{\mathrm{g}}$.
To leading order in $\delta V^{\,}_{\mathrm{g}}$, the topological gap
(\ref{eq: def topological nw gap}) changes by
$\Delta^{\,}_{\mathrm{nw}}\to\Delta^{\,}_{\mathrm{nw}}+\delta\Delta^{\,}_{\mathrm{nw}}$
with
\begin{align}
\delta\Delta^{\,}_{\mathrm{nw}}\to
-
\frac{V^{\,}_{\mathrm{g}}}{\sqrt{\Delta^{2}_{\mathrm{sc}}+V^{2}_{\mathrm{g}}}}\,
\delta V^{\,}_{\mathrm{g}}.
\end{align}
Substituting this expression into
\eqref{eq: mu estimate}
and expanding to leading order in $\delta V^{\,}_{\mathrm{g}}$,
we obtain ${{t}}\to{{t}}+\delta{{t}}$, where
\begin{align}
\frac{\delta{{t}}}{{{t}}}\approx
\frac{
\kappa-1 
     }
     {
\kappa
     }\,
\delta\kappa,
\qquad
\delta\kappa\:=
\frac{
\mathfrak{a}
     }
     {
\hbar\,v^{\,}_{\mathrm{F},\mathrm{nw}}
     }\,
\frac{V^{2}_{\mathrm{g}}}{\sqrt{\Delta^{2}_{\mathrm{sc}}+V^{2}_{\mathrm{g}}}}\,
\frac{\delta V^{\,}_{\mathrm{g}}}{V^{\,}_{\mathrm{g}}}.
\label{eq: def kappa bis}
\end{align} 
When expressed in units of the uniform hopping amplitude ${{t}}$,
we arrive at the final expressions
\begin{subequations}
\label{eq: expressing kekule gap in terms Majorana nanowire}
\begin{equation}
\frac{\delta{{t}}}{{{t}}}\approx
\frac{\kappa-1}{\kappa}\,
\frac{\mathfrak{a}}{\xi^{\,}_{\mathrm{sc}}}
\frac{
V^{2}_{\mathrm{g}}/\Delta^{2}_{\mathrm{sc}}
     }
     {
\sqrt{1+V^{2}_{\mathrm{g}}/\Delta^{2}_{\mathrm{sc}}}
     }
\frac{\delta V^{\,}_{\mathrm{g}}}{V^{\,}_{\mathrm{g}}}
\label{eq: expressing kekule gap in terms Majorana nanowire a}
\end{equation}
for the Kekul\'e perturbation
(\ref{eq: first def kekule dimerization})
with the non-uniform hopping amplitude $\delta{{t}}$, 
\begin{equation}
\frac{\Delta^{\,}_{0}}{{{t}}}\sim\frac{\delta{{t}}}{{{t}}}
\label{eq: expressing kekule gap in terms Majorana nanowire b}
\end{equation}
for the Kekul\'e gap in Eq.\ (\ref{eq: def Kekule perturbation}),
and
\begin{equation}
\xi^{\,}_{\mathrm{logical}}\:=
\frac{{{t}}}{\delta{{t}}}\,\mathfrak{a}
\label{eq: expressing kekule gap in terms Majorana nanowire c}
\end{equation}
\end{subequations}
for the decay length of a logical MZM.

Let us now show that a great deal of control over the size of the
logical MZMs is attainable using the same material parameters as in
current experimental setups. We focus on the InSb/Al systems reviewd
in~\cite{lutchyn2018}. The proximity induced superconducting gap is
$\Delta^{\,}_{\mathrm{sc}}\sim 0.2\,\mathrm{meV}$,
while the Fermi velocity can be estimated from the quoted range of
values of the spin-orbit coupling, i.e.,
$v^{\,}_{\mathrm{F},\mathrm{nw}}\sim 0.2 - 1.0\; \mathrm{eV}\times\AA$.
Hence, the proximity-induced superconducting correlation
length is in the range $\xi^{\,}_{\mathrm{sc}}\sim 100 - 500 \,\mathrm{nm}$.
For wires of length $\mathfrak{a}\sim 1\mu\mathrm{m}$,
one thus have ratios in the range
$\mathfrak{a}/\xi^{\,}_{\mathrm{sc}}\sim 2 - 10$. 

We proceed by choosing to work with $\kappa\approx 2$, which yields
significant overlap between the zero modes at the endpoints of the
wires (and can be selected via the magnetic field, as we clarify
below). According to Eq.\ (\ref{eq: def kappa a}),
this choice gives a hopping amplitude
$
t\sim
0.27\,\hbar\,v^{\,}_{\mathrm{F},\mathrm{nw}}/\mathfrak{a}=
0.27\,(\xi^{\,}_{\mathrm{sc}}/\mathfrak{a})\,
\Delta^{\,}_{\mathrm{sc}}\sim
0.027\,\Delta^{\,}_{\mathrm{sc}} - 0.14\,\Delta^{\,}_{\mathrm{sc}}
$.
The choice of working with $\kappa\approx 2$ corresponds to
a magnetic field such that
$
\Delta^{\,}_{\mathrm{nw}}\approx
\kappa\,
(\xi^{\,}_{\mathrm{sc}}/\mathfrak{a})\,
\Delta^{\,}_{\mathrm{sc}}\sim
0.2\,\Delta^{\,}_{\mathrm{sc}} - 1.0\,\Delta^{\,}_{\mathrm{sc}}$
according to Eq.\ (\ref{eq: def kappa a}).

With the choice of $\kappa \approx 2$, the Kekul\'e gap
(\ref{eq: expressing kekule gap in terms Majorana nanowire b})
is approximately given by
\begin{equation}
\frac{\Delta^{\,}_{0}}{{{t}}}\approx
\frac{1}{2}
\frac{\mathfrak{a}}{\xi^{\,}_{\mathrm{sc}}}\,
\frac{
V^{2}_{\mathrm{g}}/\Delta^{2}_{\mathrm{sc}}
     }
     {
\sqrt{1+V^{2}_{\mathrm{g}}/\Delta^{2}_{\mathrm{sc}}}
     }
\frac{\delta V^{\,}_{\mathrm{g}}}{V^{\,}_{\mathrm{g}}}.
\label{eq: expressing kekule gap in terms Majorana nanowire bis}
\end{equation}
The prefactor in front of $\delta
V^{\,}_{\mathrm{g}}/V^{\,}_{\mathrm{g}}$ on the right-hand side can be
chosen to be of order one by choosing the ratio
$V^{2}_{\mathrm{g}}/\Delta^{2}_{\mathrm{sc}}$ in the expression above
so as to compensate the factor
$
\mathfrak{a}/(2\xi^{\,}_{\mathrm{sc}})\sim
1.0 - 5.0
$.
(The corresponding bias $V^{\,}_{\mathrm{g}}$ should thus be of roughly the
same order as $\Delta^{\,}_{\mathrm{sc}}$.) If so, the ratio
$
\Delta^{\,}_{0}/{{t}}\approx
\delta V^{\,}_{\mathrm{g}}/V^{\,}_{\mathrm{g}}
$. Consequently, by using modulations with
$\delta V^{\,}_{\mathrm{g}}$ of the same order as
$V^{\,}_{\mathrm{g}}$, one can make the Kekul\'e gap of the order of
$t$, and hence the size of the logical MZMs as small as the length
scale of the wire size $\mathfrak{a}$.

We remark that for the scheme that we propose, the shorter the wires
the larger the energy scales of the effective model. The hopping
amplitude $t$ would roughly double (if one chooses to operate at the
same $\kappa\approx 2$) if one uses wires that are half as long.
(This energy scale is set by
$\hbar\,v^{\,}_{\mathrm{F},\mathrm{nw}}/\mathfrak{a}$.)
So for a 500 nm (300 nm) wire, the energy scale of
$
t\sim
0.054\,\Delta^{\,}_{\mathrm{sc}} - 0.27\,\Delta^{\,}_{\mathrm{sc}}
$
($
t\sim 0.09\,\Delta^{\,}_{\mathrm{sc}} - 0.45\,\Delta^{\,}_{\mathrm{sc}}
$)
follows.

One potential cause for concern about the hexagonal network geometry depicted in Fig.~\ref{fig:hierarchical} is the magnetic field alignment: the standard models for the low-energy physics in proximitized nanowires require a component of the applied magnetic field to be perpendicular to the spin-orbit coupling vector \cite{jason, lutchyn,oreg}, which may be problematic to achieve in a hexagonal network.
However, while the honeycomb-lattice arrangement of the nanowires depicted in Fig.~\ref{fig:hierarchical} simplifies our theoretical calculations and makes the idea transparent, this geometry is not strictly necessary in reality. For example, to simplify the magnetic field alignment in an experimental setup, one could deform the lattice into a ``brick wall'' structure, with all nanowires placed either horizontally or vertically. Then, by applying a magnetic field to the entire system at a $45^\circ$ angle, there is a nonzero component of magnetic field along each individual wire.

Another parameter relevant to experiments is the time scale on which a braiding operation can be performed such that the system remains in its ground state. One can estimate this time scale from the adiabatic theorem. The probability of transitioning to excited states when moving a single vortex a distance of a few lattice spacings can be estimated as:
\begin{eqnarray}
p_{n\neq 0} &\sim& \frac{\hbar^2}{\Delta_0^4} \langle 0 |\dot{H}^2|0\rangle_c  \nonumber \\
&\sim & \frac{\hbar^2}{\Delta_0^4} {\dot{\bm R}}^2 \left(\frac{\Delta_0}{\mathfrak{a}}\right)^2 \nonumber \\
&\sim & \frac{\hbar^2}{\Delta_0^2\mathfrak{a}^2} \dot{\bm R}^2,
\end{eqnarray}
where the dot denotes a derivative with respect to time. The adiabatic condition requires $p_{n\neq 0} \ll 1$, which leads to $|\dot{\bm R}| \ll \Delta_0\mathfrak{a}/\hbar$. Physically this means that the rate at which the vortex cores can be moved in an experiment is limited by the Kekul\'e gap.

\section{Summary}
\label{sec: summary}

In this paper, we presented a hierarchical architecture for building
“logical” Majorana zero modes using “physical” Majorana zero modes at
the Y-junctions of a hexagonal network of semiconductor nanowires. In
a nutshell, the essence of our approach is that one can build
Majoranas out of Majoranas that are, in turn, built of Majoranas (see
Fig.~\ref{fig:hierarchical}). The ``emergent'' or ``logical'' Majoranas
can be moved adiabatically and are not restricted to be centered at
sites of a lattice, although their microscopic or ``physical''
constituents are. What this construction provides is the ability to
program where one wants to place the ``logical" Majoranas by
controlling applied gate biases on the nanowires within the hexagonal
network. We present in Eq.~\eqref{eq:voltage} a simple expression for the bias voltages that
would place $v$ Majoranas at the centers of Kekul\'e vortices at
locations $\bm{R}^{\,}_{n}(t), n=1,\cdots,v$,
which can be varied as functions of time in a prescribed way.

Within the hierarchical construction of quantum Hall states, novel
quasiparticles appear as a result of condensation of other types of
quasiparticles. Such a hierarchy can be viewed within the broader
context of emergence, where novel excitations appear at different
scales. Our scheme is a form of \emph{engineered} emergence, 
where one can, by design, create novel excitations starting from
simple building blocks. In our case, we have a meta-circular
realization of Majoranas, for the emergent particles at the top of the
hierarchy coincide with those used as building blocks (those at the
bottom level of the hierarchy). The distinction between the Majoranas
at the different levels of the hierarchy is the fact that the ones on
top are movable, while the ones on the bottom are static. This is an
important difference, as the ability to move the Majoranas in the plane
in a programmable way should permit one to braid them,
providing a \textit{direct} means to probe their non-Abelian statistics.

\section*{Acknowledgment}

This work was supported by DOE Grant No. DE-FG02-06ER46316 (C.~C. and
Z.-C.~Y.) and by the Simons Foundation (C.~C.).
C.~C. acknowledges the hospitality of the Pauli Center for Theoretical
Studies at ETH Z\"urich and the University of Z\"urich, where part of
this work was carried out. T.~I. acknowledges support from the Laboratory for
Physical Sciences, and a JQI postdoctoral fellowship.
C.~M. and T.~I. acknowledge support from the
Condensed Matter Theory Visitors Program at Boston University, where part of
this work was carried out.

\bibliography{manuscript}

\end{document}